\documentclass{emulateapj}
\usepackage{natbib}

\shorttitle{Gravitational-wave Background Limits}
\shortauthors{P.~B.~Demorest et al.}

\newcommand{\us}{$\mu$s}
\newcommand{\m}[1]{\mathbf{#1}}
\newcommand{\ev}[1]{E\left\{{#1}\right\}}

\begin{document}

\title{Limits on the Stochastic Gravitational Wave Background from
the North American Nanohertz Observatory for Gravitational Waves}
\author{
P.~B.~Demorest\altaffilmark{1}, 
R.~D.~Ferdman\altaffilmark{2}, 
M.~E.~Gonzalez\altaffilmark{3}, 
D.~Nice\altaffilmark{4}, 
S.~Ransom\altaffilmark{1}, 
I.~H.~Stairs\altaffilmark{3}, 
Z.~Arzoumanian\altaffilmark{5}, 
A.~Brazier\altaffilmark{6}, 
S.~Burke-Spolaor\altaffilmark{11},
S.~J.~Chamberlin\altaffilmark{15},
J.~M.~Cordes\altaffilmark{6}, 
J.~Ellis\altaffilmark{15},
L.~S.~Finn\altaffilmark{7}, 
P.~Freire\altaffilmark{8}, 
S.~Giampanis\altaffilmark{15},
F.~Jenet\altaffilmark{9}, 
V.~M. Kaspi\altaffilmark{10}, 
J.~Lazio\altaffilmark{11}, 
A.~N.~Lommen\altaffilmark{12}, 
M.~McLaughlin\altaffilmark{13}, 
N.~Palliyaguru\altaffilmark{13},
D.~Perrodin\altaffilmark{12}, 
R.~M.~Shannon\altaffilmark{14}, 
X.~Siemens\altaffilmark{15}, 
D.~Stinebring\altaffilmark{16},
J.~Swiggum\altaffilmark{13},
W.~W.~Zhu\altaffilmark{3}}

\altaffiltext{1}{National Radio Astronomy Observatory, Charlottesville,
VA}
\altaffiltext{2}{Jodrell Bank Centre for Astrophyics, University of
Manchester}
\altaffiltext{3}{Department of Physics and Astronomy, University of British Columbia, Vacouver, BC, Canada}
\altaffiltext{4}{Lafayette College}
\altaffiltext{5}{NASA GSFC}
\altaffiltext{6}{Cornell University}
\altaffiltext{7}{Pennsylvania State University}
\altaffiltext{8}{Max Planck Institute for Radio Astronomy}
\altaffiltext{9}{University of Texas, Brownsville}
\altaffiltext{10}{McGill University}
\altaffiltext{11}{Jet Propulsion Laboratory, California Institute of
Technology}
\altaffiltext{12}{Franklin and Marshall College}
\altaffiltext{13}{West Virginia University}
\altaffiltext{14}{CSIRO Astronomy and Space Science}
\altaffiltext{15}{Center for Gravitation, Cosmology and Astrophysics,
Department of Physics, University of Wisconsin--Milwaukee}
\altaffiltext{16}{Oberlin College}

\begin{abstract}

Here we present an analysis of high-precision pulsar timing data taken
as part of the North American Nanohertz Observatory for Gravitational
waves (NANOGrav) project.  We have observed 17 pulsars for a span of
roughly five years using the Green Bank and Arecibo radio telescopes.
We analyze these data using standard pulsar timing models, with the
addition of time-variable dispersion measure and frequency-variable
pulse shape terms.  Sub-microsecond timing residuals are obtained in
nearly all cases, and the best root-mean-square timing residuals in this
set are $\sim$30--50~ns.  We present methods for analyzing post-fit timing
residuals for the presence of a gravitational wave signal with a
specified spectral shape.  These properly and optimally take into
account the timing fluctuation power removed by the model fit, and can
be applied to either data from a single pulsar, or to a set of pulsars
to detect a correlated signal.  We apply these methods to our dataset to
set an upper limit on the strength of the nHz-frequency stochastic
supermassive black hole gravitational wave background of $h_c
(1~\mathrm{yr}^{-1}) < 7\times10^{-15}$ (95\%).  This result is
dominated by the timing of the two best pulsars in the set, PSRs
J1713$+$0747 and J1909$-$3744.

\end{abstract}
\keywords{pulsars, gravitational waves, \ldots}

\section{Introduction}
\label{sec:intro}

The direct detection of gravitational radiation (or gravitational waves;
GW) is currently a major goal in experimental physics.  As described by
general relativity, GW are freely propagating wave solutions to
Einstein's equation, or ``ripples'' in the space-time metric.  Detecting
GW would confirm another key prediction of general relativity.  GW are
expected to be generated by nearly any configuration of accelerating
mass, although due to the weakness of gravity, large masses or high
accelerations are required to radiate significant GW.  This means that
astronomical objects are the only sources expected to produce
measureable GW, and that we can in turn use these detections to learn
about the GW sources themselves;  GW astronomy will provide an entirely
new window through which we can view the universe.  Binary systems are
expected to account for a large fraction of the detectable GW signals,
but we cannot discount ``exotic'' or unexpected sources either.

One of the best tools we have for making precise astronomical
measurements is the timing of radio pulses from spinning, magnetized
neutron stars (pulsars).  We receive a burst of radio waves once per
spin period as the lighthouse-like beamed emission sweeps by Earth.  The
pulse times of arrival can be analyzed via a model that counts every
single rotation of the star over years or even decades.  This provides
detailed information about the neutron star itself (via spindown rate
and spin irregularities), its binary orbit (via orbital Doppler and
relativistic effects), and astrometry.  Timing of double-neutron star
binary systems has already provided strong evidence for the existence of
GW, through measurements of orbital evolution induced by the generation
of GW by the binary system \citep{taylor:1913}.  The medium through
which the radio pulses travel from the pulsar to Earth also affects the
signal.   This has been used extensively to probe the ionized component
of the interstellar medium in a variety of ways
\citep[e.g.,][]{rickett:ism}.  The presence of GW along the line of sight
also will affect the pulse travel times.  This forms the basis for the
use of pulsars as gravitational wave detectors.

The influence of GW on pulsar timing, and its potential use in GW
detection, was first noted over 30 years ago \citep{sazhin, detw:gw}.  A
major step forward was provided by \citet{hellings:gw}, who showed that
a GW signal will produce {\it correlated} timing fluctucations in a set
of pulsars, a concept which came to be known as a pulsar timing array
\citep[PTA;][]{foster:pta}.  A second major advance was the discovery of
millisecond pulsars \citep{backer:1937disc} -- the high spin rate and
stability of these objects improves GW sensitivity by orders of
magnitude compared with canonical $\sim$1-second pulsars.  Pulsar timing
is most sensitive to GW with period comparable to the observational time
span, typically 1--10~years, so the GW frequency is in the nanohertz
band.  In this frequency range the brightest expected sources are
supermassive black hole binaries, which may be visible either as a
stochastic background \citep[e.g.,][]{jaffe:gw,sesana:gwb} or as
individual sources \citep{sesana:single}.  A number of previous upper
limits have been placed \citep{kaspi:1937, jenet:3c66, jenet:limit,
rutger:epta}, but no successful detection has yet been accomplished.
The topic continues to attract attention and a number of dedicated
pulsar timing array research projects are now active worldwide.

In this paper, we present new high-precision timing observations of 17
millisecond pulsars, covering a five-year time span.  These data were
obtained as part of the North American Nanohertz Observatory for
Gravitational Waves (NANOGrav)
project\footnote{\url{http://www.nanograv.org}}, using the two largest
single-dish telescopes available, Arecibo Observatory and the NRAO Green
Bank Telescope.  In \S\ref{sec:obs} we describe the observational setup
and dataset.  In \S\ref{sec:timing} we present our methods and results
for the determination of timing models.  In \S\ref{sec:gw} we present a
time-domain algorithm for measuring the cross-correlation between pairs
of post-fit residuals, and apply this to our timing data to either
detect or place limits on the stochastic gravitational wave background.
We discuss the astrophysical implications of these results in
\S\ref{sec:discuss} and summarize our findings in \S\ref{sec:conc}.

\section{Observations}
\label{sec:obs}

\begin{figure*}[tp]
\begin{center}
\plotone{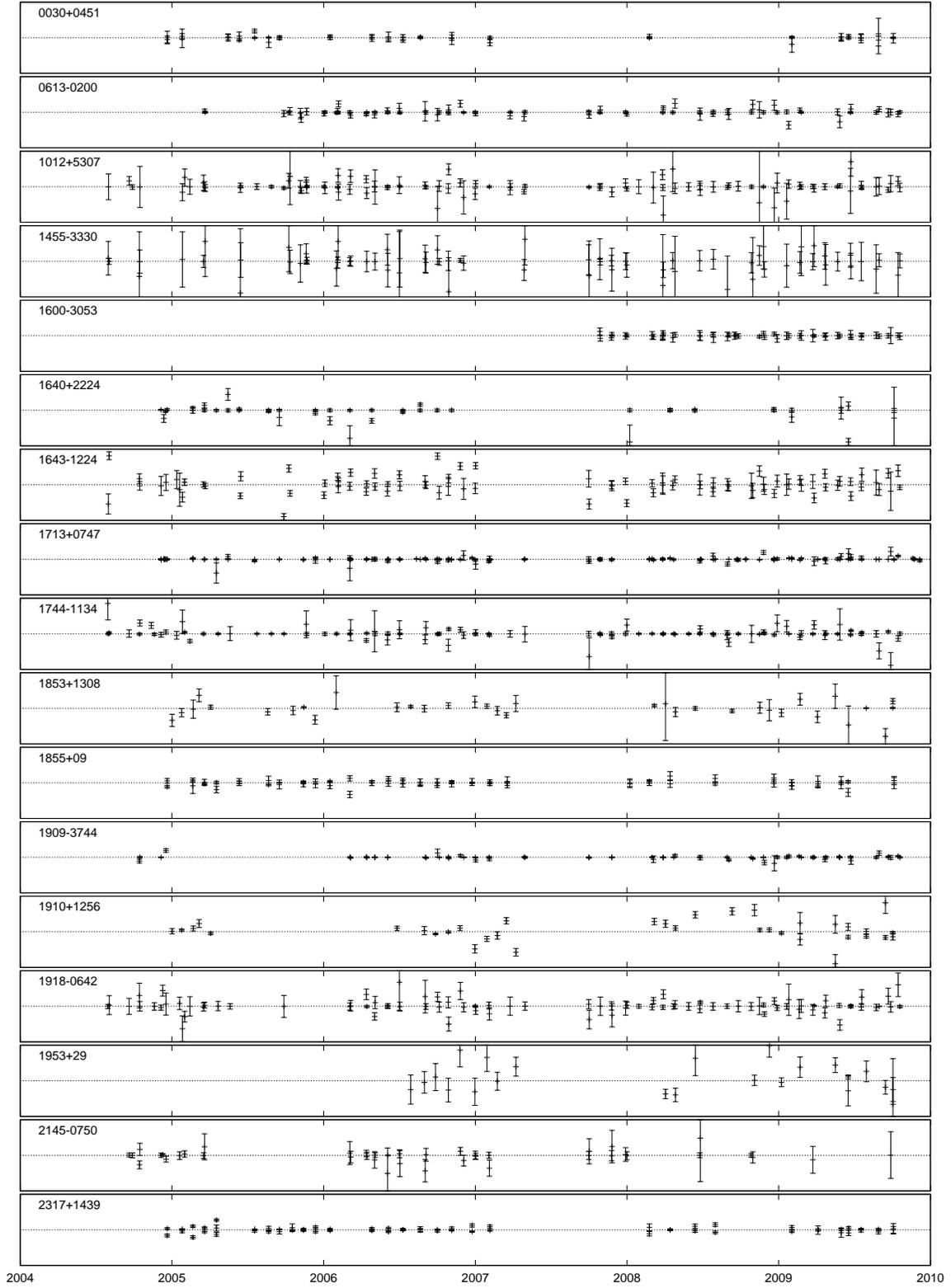}
\end{center}
\caption{\label{fig:allresids} Overview of timing residuals for all
sources, showing observational cadence and coverage during the five-year
time span.  The gap in 2007 was due to an extended maintenance
period at both telescopes.  The full scale of the y-axis is 10~\us\ in
all cases.}
\end{figure*}

The observations presented here were carried out over a five-year
period, from 2005 to 2010, as part of a program specifically designed to
measure or constrain the nHz-frequency stochastic gravitational wave
background.  The sources were observed with one or both of the 305-m
NAIC Arecibo Observatory or the 100-m NRAO Green Bank Telescope (GBT),
with a typical observational cadence of 4--6 weeks between sessions (see
Fig.~\ref{fig:allresids}).  Scheduling at each observatory was
independent, and the sessions were typically not performed
simultaneously at the two telescopes.  

These sources were selected from the known population of bright
millisecond pulsars (MSPs), excluding those found in globular clusters.
Source selection was performed with the goal of obtaining the highest
timing precision possible, to maximize sensitivity to gravitational
waves.  Due to its higher gain, Arecibo was used to observe all sources
in its visible portion of the sky ($\sim0^\circ$ to $30^\circ$
declination), while sources outside this range were observed with the
GBT, down to its minimum declination of $\sim-45^\circ$.  One source
presented in this paper, J1713$+$0747, was observed with both
telescopes.  \footnote{Another bright MSP, PSR B1937$+$21, is also
observed with both telescopes as part of this project.  We have for now
excluded it from this analysis due to its well-documented high levels of
spin noise \citep[e.g.,][]{kaspi:1937} and interstellar medium
systematics \citep[e.g.,][]{ramach:1937}.}

Each pulsar was observed at two widely separated frequencies in order to
track variation in dispersion measure (DM; see
\S\ref{sec:timing:timing}).  As the different frequency measurements use
different receivers, these observations are also not simultaneous.  The
choice of receivers used to observe a certain pulsar was determined from
the availability and performance of the receiver systems at each
telescope, and from the spectral index of the pulsar.  At Arecibo,
multi-frequency data are obtained in a single observing session, within
$\sim$1 hour for each source.  At the GBT, the different frequencies
were observed up to $\sim$1 week apart.  In a given epoch the observing
time per frequency band for each source was anywhere from 15 to 45
minutes, with shorter integration times in general at Arecibo.  The
telescopes and frequencies used to observe each source, along with its
basic physical parameters such as spin period and DM, are listed in
Table~\ref{tab:srcs}.  Three pulsars in this set (PSRs J1853$+$1308,
J1910$+$1256, and B1953$+$29) were originally observed as part of a
different project at Arecibo \citep{gonzalez:pm} and consequently only
have single-band data for most of the time span.  All receiver systems
used at 800~MHz and above are sensitive to orthogonal linear
polarizations.  The 430~MHz system at Arecibo is sensitive to dual
circular polarizations.

\begin{table*}[tbp]
\caption{\label{tab:srcs} List of observed millisecond pulsars:  Basic
parameters and observing setups.}
\begin{center}
\begin{tabular}{c|rrrr|rrrrr|c}
\hline
Source & $P$ & $dP/dt$  & DM & $P_b$ 
  & \multicolumn{5}{|c|}{Average flux density (mJy)$^a$}
  & Obs  \\
  & (ms) & ($10^{-20}$) & (pc~cm$^{-3}$) & (d) 
  &  327 MHz & 430 MHz & 820 MHz & 1.4 GHz & 2.3 GHz
  &  \\ 
\hline
J0030+0451 & 4.87 & 1.02 & 4.33 & - & - & 13.7 & - & 1.4 & - &  AO \\
J0613--0200 & 3.06 & 0.96 & 38.78 & 1.2 & - & - & 5.3 & 2.0 & - &  GBT \\
J1012+5307 & 5.26 & 1.71 & 9.02 & 0.6 & - & - & 7.6 & 3.9 & - &  GBT \\
J1455--3330 & 7.99 & 2.43 & 13.57 & 76.2 & - & - & 2.0 & 1.1 & - &  GBT \\
J1600--3053 & 3.60 & 0.95 & 52.33 & 14.3 & - & - & 3.1 & 2.3 & - &  GBT \\
J1640+2224 & 3.16 & 0.28 & 18.43 & 175.5 & - & 10.8 & - & 1.0 & - &  AO \\
J1643--1224 & 4.62 & 1.85 & 62.42 & 147.0 & - & - & 12.3 & 4.2 & - &  GBT \\
J1713+0747 & 4.57 & 0.85 & 15.99 & 67.8 & - & - & 8.8 & 6.3 & 3.6 &  AO,GBT \\
J1744--1134 & 4.07 & 0.89 & 3.14 & - & - & - & 7.6 & 2.6 & - &  GBT \\
J1853+1308 & 4.09 & 0.87 & 30.57 & 115.7 & - & - & - & 0.2 & - &  AO \\
B1855+09 & 5.36 & 1.78 & 13.30 & 12.3 & - & 14.0 & - & 4.0 & - &  AO \\
J1909--3744 & 2.95 & 1.40 & 10.39 & 1.5 & - & - & 3.4 & 1.4 & - &  GBT \\
J1910+1256 & 4.98 & 0.97 & 34.48 & 58.5 & - & - & - & 0.2 & - &  AO \\
J1918--0642 & 7.65 & 2.57 & 26.60 & 10.9 & - & - & 4.5 & 1.8 & - &  GBT \\
B1953+29 & 6.13 & 2.97 & 104.50 & 117.3 & - & - & - & 1.0 & 0.1 &  AO \\
J2145--0750 & 16.05 & 2.98 & 9.03 & 6.8 & - & - & 12.3 & 3.2 & - &  GBT \\
J2317+1439 & 3.45 & 0.24 & 21.90 & 2.5 & 32.2 & 5.4 & - & - & - &  AO \\
\hline
\end{tabular}

{$^a$ The presence of flux density values indicate the frequencies 
at which each source is observed.}

\end{center}
\end{table*}

All observations were performed using the identical Astronomical Signal
Processor (ASP) and Green Bank Astronomical Signal Processor (GASP)
pulsar backend systems \citep{demorest:phd} at Arecibo and the GBT
respectively.  These systems perform real-time coherent dedispersion,
full-Stokes detection, and pulse period folding in software, using a
$\sim$20-node Linux-based computer cluster.  Channelized voltage data
are supplied to the cluster from a SERENDIP-V FPGA
board\footnote{\url{https://casper.berkeley.edu/galfa/}} controlled by a
Compact PCI single-board computer.  The SERENDIP-V board performs
analog-to-digital conversion and a digital polyphase filterbank
operation to split a 128-MHz bandwidth, dual-polarization signal input
from the telescope into 32 4-MHz wide channels.  Both the initial
digitization of the signal and the channelized output use 8-bit complex
quantization.  Further details of the hardware and real-time software
used in these observations were described by \citet{demorest:phd}.
While the 4-MHz channelization provides a minimum time resolution of
250~ns, the final time resolution depends on the pulse period and number
of profile bins used to average the profile -- in this case, either 2048
(GBT) or 4096 (Arecibo) bins were used.  Profiles were typically
integrated for 1 minute (Arecibo) or 3 minutes (GBT).  The total
bandwidth processed varied with pulsar and observing frequency, limited
by either the real-time computational load or the receiver bandpass.  A
maximum of 64~MHz was used in most cases, with smaller bandwidths down
to $\sim$20~MHz used for low-frequency or high-DM observations.

\section{Timing Analysis}
\label{sec:timing}

In this section, we describe the procedures used in the timing portion
of the data analysis.  These can be split into two main areas:  analysis
dealing with pulse profiles, including polarization calibration and
determination of pulse times of arrival (\S\ref{sec:timing:toa}); and
fitting a physical timing model to the arrival times
(\S\ref{sec:timing:timing}).  The further analysis step of determining
GW background limits is discussed later, in \S\ref{sec:gw}.

\subsection{Calibration and Time of Arrival Estimation}
\label{sec:timing:toa}

As discussed in \S\ref{sec:obs}, the data products resulting from an
observing session are a set of full-Stokes pulse profiles integrated
over 4~MHz radio bandwidth and 1--3 minutes of time, into either 2048 or
4096 pulse phase bins.  Following standard pulsar data analysis
procedures, we aim to determine from the profile data a set of pulse
times of arrival (TOAs), i.e. times at which the apparent rotational
phase of the pulsar passes through some fiducial point.  This process
involves several major steps:  polarization calibration, template
profile creation, additional profile averaging, and finally TOA
measurement.  For this work, we have performed all of these steps using
two independent versions of pulsar data processing software, the first
based on the PSRCHIVE\footnote{\url{http://psrchive.sourceforge.net}}
software \citep{psrchive}, and the second based on ASPFitsReader
\citep{ferdman:phd}.  This procedure provides an important cross-check
for errors in the analysis software that might otherwise be hard to
detect.  Both analysis pipelines performed the same procedures, aside
from template determination which was done only via PSRCHIVE.

Pulsar radio emission is typically highly polarized.  While our current
analysis relies only on the total intensity profiles for TOA
determination, due to the high degree of polarization, calibration
errors can still distort the intensity profile shape, leading to a TOA
bias \citep[e.g.,][]{straten:mtm}.  A complete description of the
instrumental response to a polarized signal is provided by the Mueller
matrix, which is a radio-frequency-dependent linear transformation from
the intrinsic to observed Stokes parameters
\citep{heiles:pol,straten:cal}.  While we plan in future work to apply
full Mueller matrix calibration to these data, for the current analysis
we correct only the leading order terms, differential gain and phase
between the two polarization components of the telescope signal.  This
is done via an injected calibration signal; immediately before or after
each pulsar observation, a noise diode switched at 25~Hz is coupled into
both polarization signal paths and measured with the pulsar backends.
This provides a constant reference power versus frequency that is used
to scale the pulsar data.  The equivalent flux density of the
calibration signal is determined separately for each polarization by
observing the noise diode along with a bright, unpolarized quasar of
known flux density (B1442$+$101 at Green Bank, J1413$+$1509 at Arecibo).
This then provides a second scaling to convert the pulsar data to flux
density units (Jy) separately in each linear (or circular for certain
receivers) polarization.  The two calibrated total power terms are then
added together to form the total intensity (``Stokes I'') profile, and
the polarized profiles are not used further in this analysis.

\begin{figure*}[tp]
\begin{center}
\plotone{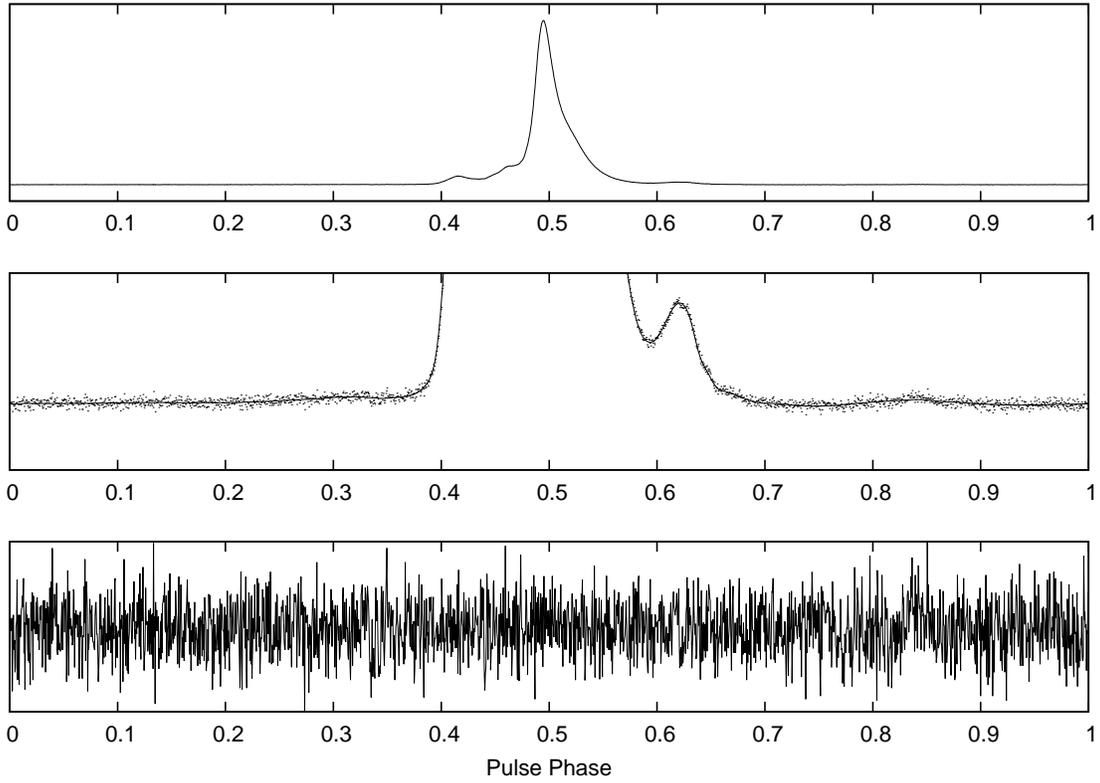}
\end{center}
\caption{\label{fig:templ} Full-sum profile and template profile for
J1713+0747 at 1400~MHz.  The top panel shows the full-sum profile at
full scale.  The middle panel shows the full-sum profile (points) and
wavelet-denoised template version (line).  The bottom panel shows the
residual difference between the full-sum profile and template.}
\end{figure*}

The calibrated pulse profiles are used to determine pulse TOAs and their
uncertainties by fitting for a pulse phase shift between each profile
and a standard ``template'' profile \citep{taylor:fftfit}.  The template
ideally is a noise-free representation of the average pulse profile
shape; any noise present in the template, especially if correlated with
noise in the profiles, can bias the TOAs \citep{hotan:0737}.  For this
work we determine template profiles from our measured data in a two-step
process:  the profiles are first roughly aligned using a single-Gaussian
template, then are summed together using weights to optimize the S/N in
the final full-sum profile \citep[see][Eqn.~2.10]{demorest:phd}.  We
then apply a translation-invariant wavelet transform and thresholding
\citep{coifman:tide} to the profile to remove its noise
(Figure~\ref{fig:templ}).  The procedure is then iterated once, now
using the new template for alignment, to produce the final template
profiles used for determining TOAs.  The wavelet noise removal procedure
is implemented in the PSRCHIVE program \verb+psrsmooth+.  In this way,
we obtained one template per pulsar per receiver used to observe it.
While all other processing steps in this section were performed
independently by both software pipelines, for consistency the same set
of templates was used in both cases.

The standard template-matching procedure used for TOA determination in
this analysis is known to suffer problems when applied to low-S/N data
\citep{hotan:0737}.  Therefore, it is useful to average profiles over as
much time and radio bandwidth as possible to maximize the S/N ratio
before forming TOAs, while still retaining enough resolution to measure
all appropriate instrumental or astrophysical effects.  In this analysis
we will retain the native instrumental frequency resolution (see
\S\ref{sec:timing:timing}), and have chosen to average over time all
profiles in a given frequency channel from each observing epoch
(typically 30 minutes).  TOAs are then measured for each average
profile, resulting in a set of $\sim$20--30 TOAs (one per 4-MHz
frequency channel) per each dual-receiver pair of observing epochs.  The
total number of TOAs for each pulsar are listed in Table~\ref{tab:fits}.
As previously mentioned, we computed TOAs using two independent analysis
pipelines.  After verifying that the two pipelines produced consistent
results, we focus the remainder of the analysis on the PSRCHIVE-produced
data; all further results presented in this paper are specific to these
data.  These TOAs are the inputs for the next part of the analysis
procedure, fitting the timing model.  The entire set of TOAs used in
this analysis can be obtained as an electronic supplement to this
paper.\footnote{\url{http://www.cv.nrao.edu/$\sim$pdemores/nanograv\_data}}

\begin{table*}[tp]
\caption{\label{tab:fits} Overview and results from timing model fits.}
\begin{center}
\begin{tabular}{c|r|rrr|rr|ccc}
\hline
Source & \# of & \multicolumn{3}{|c|}{\# of  parameters} & RMS & Fit
$\chi^2$ & \multicolumn{3}{|c}{Epoch-averaged RMS ($\mu$s)$^c$} \\
  & TOAs$^a$ & DM & Profile & Other$^b$ & ($\mu$s) &  & Low-band$^d$ & High-band &
  Combined\\
\hline
J0030+0451 & 545 & 20 & 26 & 7 & 0.604 & 1.44 & 0.019 & 0.328 & 0.148 \\
J0613--0200 & 1113 & 34 & 45 & 12 & 0.781 & 1.21 & 0.021 & 0.519 & 0.178 \\
J1012+5307 & 1678 & 52 & 53 & 14 & 1.327 & 1.40 & 0.192 & 0.345 & 0.276 \\
J1455--3330 & 1100 & 37 & 53 & 12 & 4.010 & 1.01 & 0.363 & 1.080 & 0.787 \\
J1600--3053 & 625 & 21 & 31 & 14 & 1.293 & 1.45 & 0.233 & 0.141 & 0.163 \\
J1640+2224 & 631 & 23 & 26 & 12 & 0.562 & 4.36 & 0.057 & 0.601 & 0.409 \\
J1643--1224 & 1266 & 40 & 48 & 13 & 2.892 & 2.78 & 0.589 & 1.880 & 1.467 \\
J1713+0747 & 2368 & 50 & 111 & 15 & 0.106 & 1.48 & 0.092 & 0.025 & 0.030 \\
J1744--1134 & 1617 & 54 & 49 & 7 & 0.617 & 3.58 & 0.139 & 0.229 & 0.198 \\
J1853+1308 & 497 & 0 & 34 & 12 & 1.028 & 1.16 & 0.271 & 0.096 & 0.255 \\
B1855+09 & 702 & 29 & 21 & 14 & 0.395 & 2.19 & 0.277 & 0.101 & 0.111 \\
J1909--3744 & 1001 & 31 & 37 & 14 & 0.181 & 1.95 & 0.011 & 0.047 & 0.038 \\
J1910+1256 & 525 & 0 & 34 & 14 & 1.394 & 2.09 & 0.712 & 0.684 & 0.708 \\
J1918--0642 & 1306 & 49 & 37 & 12 & 1.271 & 1.21 & 0.129 & 0.211 & 0.203 \\
B1953+29 & 208 & 0 & 27 & 12 & 3.981 & 0.98 & 1.879 & 0.543 & 1.437 \\
J2145--0750 & 675 & 20 & 37 & 12 & 1.252 & 1.97 & 0.068 & 0.494 & 0.202 \\
J2317+1439 & 458 & 30 & 12 & 15 & 0.496 & 3.03 & 0.373 & 0.150 & 0.251 \\
\hline

  \multicolumn{10}{l}
  {\footnotesize $^a$ One TOA per frequency channel per epoch.} \\

  \multicolumn{10}{l}
  {\footnotesize $^b$ ``Other'' parameters are all spin, astrometric and
  binary parameters as described in \S\ref{sec:timing:timing}.}\\

  \multicolumn{10}{l}
  {\footnotesize $^c$ RMS computed from residuals averaged down to one
  point per receiver per epoch.  See text for details.} \\

  \multicolumn{10}{l}
  {\footnotesize $^d$ Note that in these results, the low-frequency RMS
  tends to be suppressed due to the DM$(t)$ fit.}\\

\end{tabular}
\end{center}

\end{table*}

\subsection{Timing Model Fit}
\label{sec:timing:timing}

The second part of the timing analysis is to fit the measured pulse TOAs
for each pulsar to a physical timing model.  The timing model predicts
the apparent rotational phase of a pulsar based on a set of physical
parameters describing the star's rotation (spin period, spin-down rate),
astrometry (position, proper motion, parallax), binary orbital motion,
and general relativistic effects such as Shapiro delay.  The model
prediction is compared to the measured TOAs and best-fit parameter
values are determined via $\chi^2$ minimization.  This procedure is a
fundamental part of pulsar astronomy and has been described many times
in the literature \citep[see for example][]{lorimer:book}.  For all
results presented here, we use the standard
TEMPO\footnote{\url{http://tempo.sourceforge.net}} timing analysis
software.  Preliminary comparisons with analysis using the newer
TEMPO2\footnote{\url{http://tempo2.sourceforge.net}} package
\citep{hobbs:tempo2} produce nearly identical results and will be
presented in future works \citep{perrodin:noise,ellis:cov}.

As the main goal of this analysis is to detect or limit the nHz
gravitational wave background, a detailed discussion of each pulsar's
timing model parameters and the astrophysical significance of the
results will not be presented here.  The TEMPO parameter files
containing the model parameters and fit results can be obtained along
with the TOAs in the electronic supplement to the paper.  However, the
overall strategy for fitting the timing models can be described as
follows:
\begin{enumerate}

  \item{The average pulsar spin frequency and frequency derivative
  (spin-down) were always fit parameters.  No higher frequency
  derivatives were included.}

  \item{All five astrometric terms -- two sky coordinates, two
  components of proper motion, and parallax -- were always fitted
  parameters.}

  \item{Binary systems were fit for all five Keplerian parameters, using
  ether the ``ELL1'' or ``DD'' \citep{binary:dd} timing models as
  appropriate.  Additional relativistic or secular orbital terms were
  added only if the timing was significantly improved.}

\end{enumerate}

\begin{figure*}[tp]
\begin{center}
\plotone{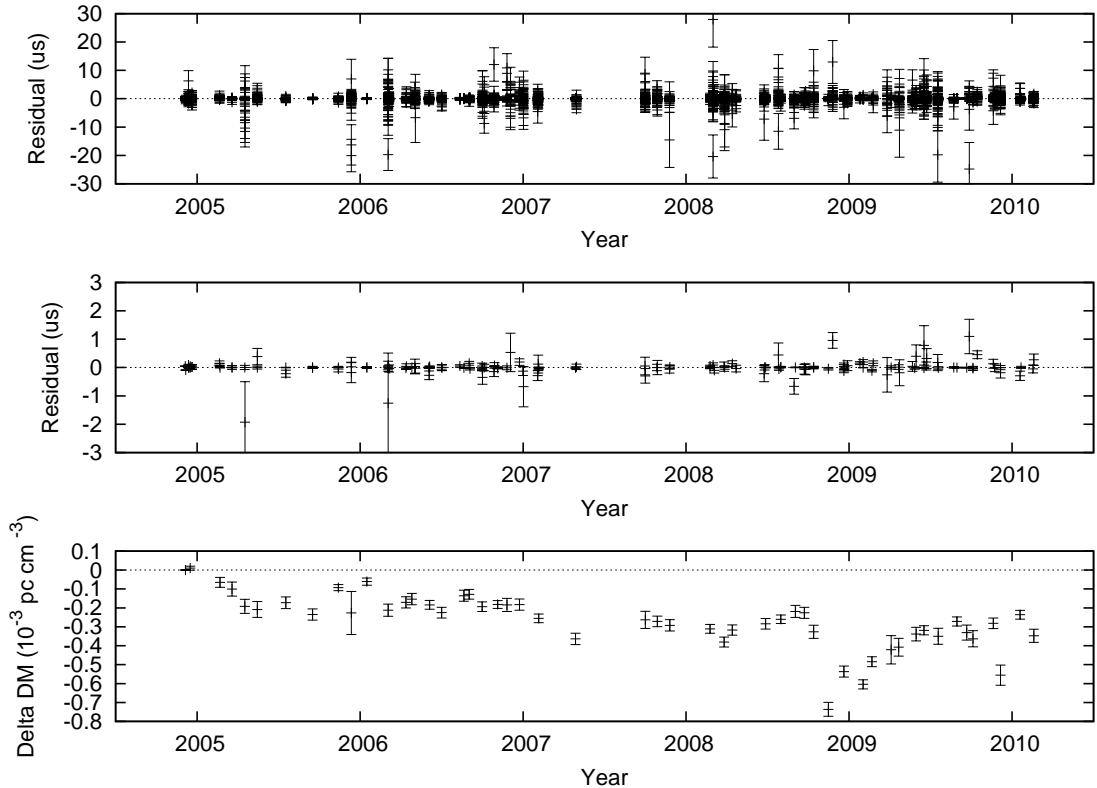}
\end{center}
\caption{\label{fig:1713:timing}
Timing summary for PSR J1713+0747.  The top panel shows residuals from the
multi-frequency TOAs used in the timing fit.  The middle panel shows the
same residuals averaged down to one point per band per day.  The bottom
panel shows the measured variation in DM as a function of time.}
\end{figure*}

In addition to the model parameters just described, the timing fit also
included terms to correct for time-variable DM and frequency dependence
of the pulse profile shape.  Dispersion is a propagation effect caused
by the travel of the radio pulses through the ionized interstellar
medium (ISM).  This causes a radio frequency-dependent shift in pulse
arrival time proportional to $\mathrm{DM} \nu^{-2}$.  Astrophysically,
DM represents the integrated column density of free electrons along the
line of sight to the pulsar.  Due to relative motion of the Earth and
the pulsar, the effective path through the ISM changes with time, hence
the DM varies in a stochastic manner \citep[e.g.][]{ramach:1937}.  For
this experiment, we obtained (non-simultaneous) dual-frequency data as
described in \S\ref{sec:obs} specifically to measure and remove this
effect from the timing.  We have done this by including a
piecewise-constant $\mathrm{DM}(t)$ function in the same fit that
determines the other timing model parameters.  For each observing epoch,
a window of span up to 15 days is defined over which an independent DM
is fit for.  Epochs for which no dual-frequency data exist within a 15
day range were excluded from the analysis.  This leads to a DM versus
time measurement for each pulsar, as shown in
Figure~\ref{fig:1713:timing}.  As previously noted, PSRs J1853$+$1308,
J1910$+$1256 and B1953$+$29 have mostly single-band observations, so the
DM variation has not been modelled for these three sources.

During the course of this analysis, we discovered additional radio
frequency-dependent trends in pulse arrival times that were not well
described by the $\nu^{-2}$ dispersion relation.  We attribute these to
the intrinsic evolution of the pulse profile shape with frequency
\citep[e.g.][]{kramer:spectra}.  The interstellar medium is another
possible cause, however since the effect does not appear to be
time-variable an ISM explanation seems less likely.  Regardless of its
physical origin, any profile shape change versus frequency can lead to
systematic TOA biases as follows: Although we have used separate
template profiles per receiver, within each receiver band the template
profile is constant.  If the true profile shape is changing versus
frequency, this will cause small but measureable systematic effects.  To
correct for these biases, we have included as free parameters in the
timing fit a constant (in time) offset -- also known as a ``jump'' --
for each 4~MHz frequency channel.

A summary of the timing analysis and results is presented in
Table~\ref{tab:fits}.  For each pulsar, the total number of TOAs, and
the total number of fit parameters is given.  These are divided into
those relating to the DM variation, frequency-dependent terms, and the
other standard spin, astrometric and binary parameters.  The success of
the timing model fit can be characterized by analyzing the post-fit
residuals, i.e. the difference between the observed and model-predicted
arrival times.  The uncertainty-weighted root-mean-square (RMS) residual
value and normalized $\chi^2$ values as determined directly from the fit
are listed in the table.  The majority of $\chi^2$ values fall near 1,
but are as high as $\sim$4 in some cases.  These values have been
computed without the use of any multiplicative or additive modifications
to the TOA uncertainties.\footnote{I.e., the ``EFAC'' and ``EQUAD''
options respectively in TEMPO.  These parameters were sometimes used in
previous timing analyses to compensate for unexplained systematic errors
in TOAs or their uncertainties.} Traditionally pulsar timing results
have been presented using TOAs which come from an average of all data
taken during a given day.  Here, our fit procedure requires the
multifrequency data be kept separate, so we cannot perform this
averaging pre-fit.  To facilitate comparison with previous work, the
final three columns in Table~\ref{tab:fits} show the ``epoch-averaged
RMS'' computed by first averaging the raw post-fit {\it residuals} down
to one point per receiver per epoch, then taking their weighted RMS.
The RMS from the low-frequency and high-frequency bands for each pulsar
are shown separately as well as the combined value using all the data.
Due to the DM$(t)$ fit, the low-frequency RMS tends to be suppressed,
making it less useful as a simple characterization of the timing.  The
two best pulsars in the set, PSRs J1713$+$0747 and J1909$-$3744, both
have epoch-averaged RMS in the $\sim$30--50~ns range with comparable
values for both the high-frequency and combined versions of this
statistic.

\section{Gravitational Wave Analysis}
\label{sec:gw}

The presence of gravitational waves (GW) along the line-of-sight from a
pulsar to Earth alters the effective path length in a time-variable
manner, resulting in extra perturbations in pulsar timing residuals
\citep{detw:gw}.  The observed amplitude of a single pulsar's timing
residuals can therefore be used to limit the strength of any GW that may
exist \citep{kaspi:1937,jenet:limit}.  However, as an observed timing
perturbation could also arise from a number of non-GW sources, this
method can not be used to definitively confirm the presence of GW.  It
was first noted by \citet{hellings:gw} that a GW signal induces {\it
correlated} variations in the timing residuals of a set of pulsars.  The
form of this correlation is unique to GW among the expected
perturbations, and attempting to detect it is the basis for most current
pulsar timing array (PTA) efforts \citep{rutger:epta,yardley:ppta},
including this work.

In this analysis, we will search for a stochastic gravitational wave
background (GWB) signal in the pulsar timing results presented in
\S\ref{sec:timing}.  We assume the GWB will take the form which has
become standard in this field -- a power-law frequency spectrum and
isotropic angular distribution.  This signal is expected to be generated
by the sum of unresolved supermassive black hole binary systems with
masses of $\sim10^8$~M$_\odot$ and orbital periods of 1--10~years.  In
this case, the characteristic strain spectrum is expected to have a
``red'' power-law spectral index $\alpha = -2/3$
\citep[e.g.,][]{jaffe:gw,sesana:gwb}, with a possible break near 10~nHz
due to the finite number of sources \citep{sesana:gwb}.  A GWB of this
form could also be generated from cosmic superstrings, with $\alpha =
-7/6$ \citep{damour:string,siemens:string}, or as inflationary
``relics'' with spectral index $\alpha = -1$ \citep{grishchuk:relic}.
Any GW signal will produces correlation in the timing fluctuations of
pairs of pulsars.  In the specific case of an isotropic GWB, the amount
of correlated power is a function only of the angular separation of the
two pulsars in the pair, and has a characteristic functional form first
predicted by \citet{hellings:gw}.

In this paper, we adopt definitions of the expected gravitational wave
spectrum and its effect on timing consistent with previous papers on the
topic \citep[e.g.,][]{jenet:limit,rutger:epta}.  In particular, we
assume a power-law spectrum in characteristic strain,
\begin{equation}
\label{eqn:hc}
  h_c(f) = A_{f_0} \left( \frac{f}{f_0} \right)^\alpha.
\end{equation}
Here, $A_{f_0}$ is the unknown GW spectrum amplitude at a reference
frequency $f_0$.  For consistency with previous literature, we set $f_0
= 1~\mathrm{yr}^{-1}$, and will call the resulting amplitude $A_1$.
This GWB produces a fluctuation $y(t)$ in the pulse times of arrival
from a given pulsar, with power spectrum given by
\begin{equation}
\label{eqn:ps_timing}
  S_y(f) = \frac{1}{12\pi^2} \frac{1}{f^3} h_c(f)^2.
\end{equation}
It is important to note that $y(t)$ represents the {\it pre-fit}
contribution of the GW signal to the pulse TOAs.  The effect of the
timing model fit will be considered in \S\ref{sec:gw:timing}.  Also
important is that this formulation of $S_y(f)$ is consistent with that
used by \citet{jenet:limit}, \citet{hobbs:tempo2_gw}, and
\citet{rutger:epta}\footnote{While the equations given by
\citet{rutger:epta} used the \citet{jenet:gw} definition, their
published limit of $A_1<6\times10^{-15}$ was computed using the same
scaling as our Eqn.~\ref{eqn:ps_timing} (van Haasteren 2012, private
communication).} but is a factor of $3$ smaller than that used by
\citet{jenet:gw}.  This results in a factor of $\sqrt{3}$ difference in
limits on $A_1$ depending on which definition is in use, and care should
therefore be taken when directly comparing the various published limits.

It is useful to compute the expected time-domain correlation between
pairs of timing fluctuations from pulsars $a$ and $b$,
\begin{equation}
\label{eqn:gw_cov}
  C_{y,ij}^{(ab)} = \ev{y_a(t_i) y_b(t_j)}
    = C_y(t_i - t_j) \zeta(\theta_{ab}).
\end{equation}
Here, $\theta_{ab}$ is the angular separation between pulsars $a$ and
$b$, and $\zeta(\theta_{ab})$ is the Hellings-Downs function describing
the expected angular correlation for a isotropic GWB.  $C_y(\tau)$ is
the GW signal autocorrelation (Fourier transform of $S_y(f)$), and can
be computed analytically for a power-law spectrum as presented by
\citet{rutger:gw, rutger:epta}.  Additionally, here and in the
following, $\ev{\cdot}$ represents the statistical expectation value, or
average over many realizations of the enclosed quantity.  This
description assumes the GW signal follows wide-sense stationary
statistics \citep[e.g.,][]{papoulis}.

In the following sections we first present methodology for analyzing the
timing residuals from \S\ref{sec:timing}, and discuss the effect that
fitting the timing model has on the statistics of post-fit residuals.
We will then apply these methods to first determine an upper limit to
the stochastic GWB amplitude using a single pulsar, and then attempt to
detect or limit the GWB by looking for angular correlations.

\subsection{Effect of the Timing Model Fit}
\label{sec:gw:timing}

In order to use pulsar timing data to detect gravitational waves, one
must account for the fact that the pulsar parameters (spin period,
astrometric parameters, etc) are not known {\it a priori}, and need to
be determined from the same data that are used for GW detection.
Especially in the case of red GW spectra, a large fraction of the GW
signal power is covariant with the ``long-term'' pulsar parameters such
as spin period and spin-down rate, and cannot be unambiguously separated
from these intrinsic pulsar features.  Previous analyses have typically
dealt with this by decomposing timing residuals using a set of
polynomials of cubic and higher order \citep{kaspi:1937, jenet:limit}.
These are approximately orthogonal to the rest of the timing fit, and
capture most of the low-frequency GW power.  Recent analyses by
\citet{rutger:gw, rutger:epta} use a Bayesian framework to limit or
detect GW signals in the data, marginalizing over the timing model
parameters.  \citet{yardley:ppta} use an extensive series of simulations
to characterize the effect of timing parameter fitting on the GW signal
in the frequency domain.

In this work, we perform the GW analysis in a separate step from the
timing fit, while using the properties of the timing fit to both
determine how much GW power is absorbed by it, and how to best detect or
limit that which remains.  This is built around a computation of the
statistics (covariance matrix) of the post-fit residuals.  This analysis
draws heavily on that presented by \citet{demorest:phd}, which can be
referred to for additional discussion.  Further detailed discussion and
application of this method will be presented by \citet{ellis:cov}.  It
is also interesting to note that a very similar calculation appears in
the Bayesian method presented by \citet{rutger:gw}, as a means of
marginalizing over the timing fit ``nuisance'' parameters.  

To begin we note that the fit that determines the pulsar
parameters is a weighted least-squares minimization, which in matrix
terms is the solution of the normal equations \citep[see for
example][]{nr}:
\begin{equation}
\label{eqn:lsfit}
 \m{A}^T \m{W} \m{A} \m{a} = \m{A}^T \m{W} \m{y}
\end{equation}
Here $\m{y}$ is the vector of input data (sometimes called ``pre-fit
residuals''), $\m{A}$ is the fit design matrix whose columns are the fit
basis functions, $\m{a}$ is the vector of fit parameter values, and
$\m{W}$ is the weighting matrix.  In a standard weighted least squares
fit such as is used in this paper, $\m{W}$ is a diagonal matrix whose
entries are calculated from the TOA uncertainties, $W_{ii} =
\sigma_i^{-2}$.  However, we write Eqn.~\ref{eqn:lsfit} in this general
form to emphasize that the following methods are equally valid for a
generalized least-squares fit in which $\m{W}$ can have non-zero
off-diagonal elements.  The generalized approach has recently been
proposed for pulsar timing analyses as a way of handling correlated
noise in the timing data \citep{coles:gls}.  In the generalized case,
$\m{W}$ is the inverse of a non-diagonal noise covariance matrix.

The post-fit residuals $\m{r} = \m{y} - \m{A}\m{a} = \m{R}\m{y}$ are
therefore a linear function of the input data, and can be determined by
applying the residual projection operator $\m{R}$:
\begin{equation}
  \m{R} = \m{I} - \m{A}\left(\m{A}^T\m{W}\m{A}\right)^{-1} 
      \m{A}^T \m{W}
\end{equation}
A key feature of writing the analysis in this form is that althought
$\m{R}$ does depend of the weighting matrix $\m{W}$, it does not depend
on the data values $\m{y}$.  Therefore, for a given pattern of weights
(or uncertainties), the action of the fit can be studied independent of
a specific data realization.  This feature can be used to eliminate
dependence on simulation in interpreting results and determining GW
limits, and can provide additional insight into the properties of the
fit procedure.  In particular, given a known or assumed input data
covariance matrix $\m{C_y}$, the resulting residual covariance matrix
$\m{C_r}$ can be easily calculated:
\begin{equation}
\label{eqn:r_cov}
  \m{C_r} = \ev{\m{r}\m{r}^T} = \m{R} \ev{\m{y}\m{y}^T} \m{R}^T
    = \m{R} \m{C_y} \m{R}^T
\end{equation}
The expected cross-covariance between the timing residuals of a pair of
pulsars ($a$ and $b$) can be calculated similarly:
\begin{equation}
\label{eqn:r_xcov}
  \m{C_{r_a,r_b}} = \ev{\m{r_a} \m{r_b}^T} 
    = \m{R_a} \m{C_{y_a,y_b}} \m{R_b}^T
\end{equation}

It is worth noting that the ensemble of post-fit residuals is in general
non-stationary.  That is, the elements of the covariance matrix,
$C_{r,ij}$, are not simply a function of the time lag
$\tau_{ij}=t_j-t_i$, even if the pre-fit stochastic process represented
by $\m{C_y}$ {\it does} have this property.  Therefore the action of the
timing fit cannot be fully represented as a filter or frequency domain
transfer function, as has sometimes been proposed in the past
\citep[e.g.,][]{blandford:timing,jenet:gw}.  This feature of the
statistics of the residuals, along with the irregular sampling, varying
data span and quality, and presence of steep-spectrum red noise
processes, present significant obstacles to a frequency domain analysis
of real-world pulsar timing data.  This motivates our decision to
perform the GW analysis in the time domain.

\subsection{Single-pulsar Analysis}
\label{sec:gw:single}

We can use the methods discussed in the previous section, along with
timing data from a single pulsar, to compute upper limits on the
strength of the GWB.  As previously discussed, there is no definitive
way to distinguish GW-induced timing fluctuations observed in a single
pulsar from those due to intrinsic pulsar spin irregularity or other
non-GW effects.  However, in the low-amplitude GW regime, focusing on
the single best pulsar in this manner can provide some of the most
constraining upper limits.

\begin{figure*}[tp]
\begin{center}
\plotone{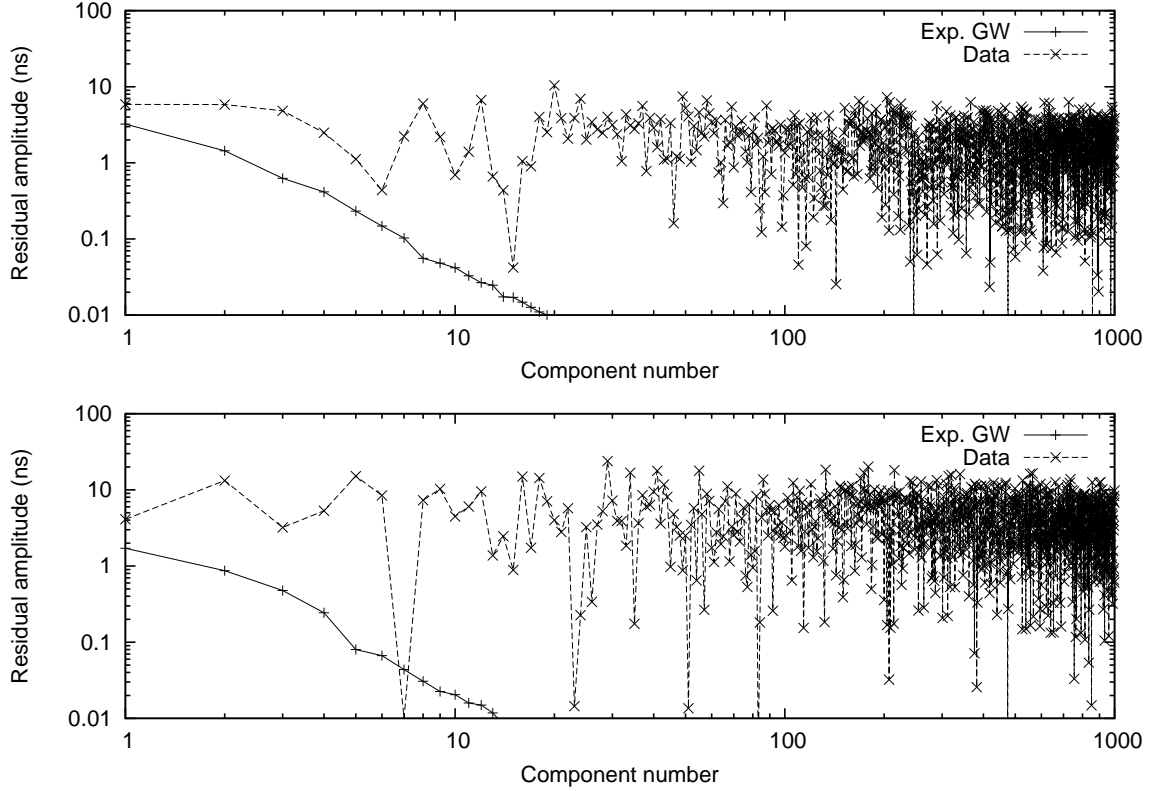}
\end{center}
\caption{\label{fig:evals}
Eigenvalue spectra for J1713$+$0747 (top) and J1909$-$3744 (bottom).  In
each plot, the line labelled ``Data'' shows the absolute value of the
residual coefficients $c_i$ when transformed into the optimal basis for
an $\alpha=-2/3$ power-law GW spectrum.  The ``Exp. GW'' line shows the
expected amplitude (square root of the eigenvalues, $\lambda_i^{1/2}$)
for an $A_1=10^{-15}$ GW signal level.  The fact that the data lines are
flat reflects the fact that the residuals are dominated by white noise.
}
\end{figure*}

Given a GW spectrum of assumed power-law shape but unknown amplitude, we
compute the pre-fit GW covariance matrix $\m{C_y^{GW}}$ up to an overall
scaling by $A_1^2$ (see Eqn.~\ref{eqn:gw_cov} and \citet{rutger:gw}).
This is then converted to the residual covariance matrix $\m{C_r^{GW}}$
using Eqn.~\ref{eqn:r_cov}.  Diagonalizing $\m{C_r^{GW}}$ provides an
orthonormal basis of eigenvectors that can be used to decompose the
timing residuals.  This basis represents the GW signal using the
smallest possible number of components,\footnote{This is very similar to
prinicpal components analysis (PCA).  The only difference is that PCA is
typically based on an empirical data covariance matrix rather than an
assumed covariance matrix as we have used here.} and is completely
orthogonal to the timing fit.  In other words, this basis optimally
captures the portion of the GW signal in the data that is not absorbed
by the timing model fit.  Transforming the residuals into this basis
produces a set of projection coefficients $c_i$.  The associated
eigenvalues ($\lambda_i$) give the relative level of GW signal power in
each component, and by assuming a nominal value of $A_1$, can be scaled
to produce expected mean-square timing residual values.
Figure~\ref{fig:evals} shows the measured and expected component
amplitudes for the two best pulsars in our set, PSRs J1713$+$0747 and
J1909$-$3744, assuming $\alpha=-2/3$ and $A_1=10^{-15}$.  In these plots
and the following discussion, the residual coefficients and eigenvalues
have been normalized to represent the RMS residual due to each
component.  The quadrature sum of all coefficients for a given pulsar
$\left( \sum c_i^2 \right)^{1/2}$ reproduces its full RMS residual as
presented in Table~\ref{tab:fits}.

\begin{figure*}[tp]
\begin{center}
\plotone{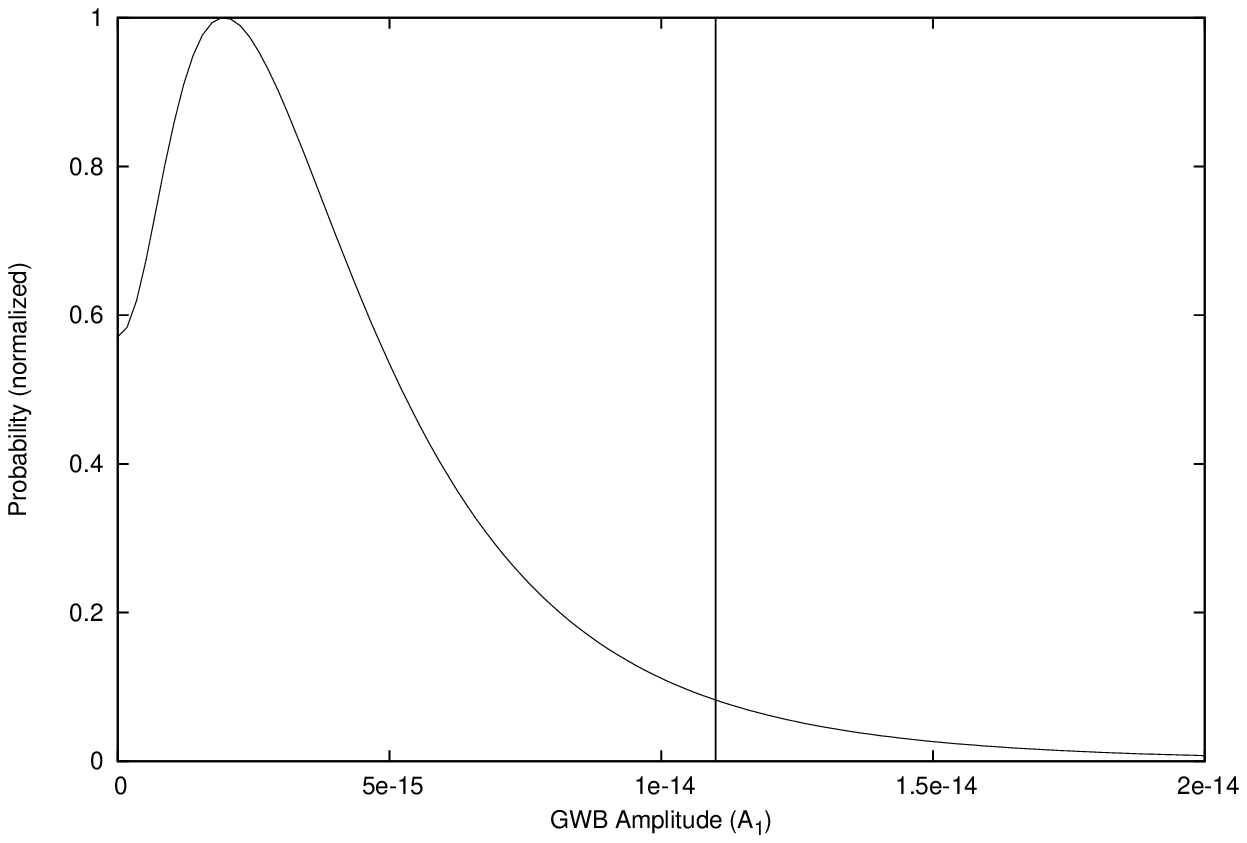}
\end{center}
\caption{\label{fig:amp_pdf}
Probability distribution for the GWB amplitude $A_1$ based on
J1713$+$0747 timing, and assuming $\alpha=-2/3$.  95\% of the
distribution is contained in $A_1 < 1.1\times10^{-14}$ (vertical line),
the maximum-likelihood value $\hat{A_1}=1.9\times10^{-15}$, and the
likelihood ratio between this point and $A_1=0$ is $R=0.6$.  In this
case, a non-zero value of $A_1$ provides the best fit to the data, but
without much statstical significance over a white noise only model.
These statistics for all sources, considering several different values
of $\alpha$, are presented in Table~\ref{tab:apdf}.
}
\end{figure*}

It is immediately apparent that the data shown in a plot such as
Figure~\ref{fig:evals} can be used to place a limit on the strength of a
power-law GWB.  Roughly speaking, any GW signal present in the data must
lie near or below the observed data values.  For measured component
values $c_i$ and eigenvalues $\lambda_i$, we combine several components
into a single limit by forming the likelihood function for the
parameters $A_1$ and $\sigma$, the level of white noise present in the
data:
\begin{equation}
\label{eqn:logl}
\log L(A_1, \sigma) = -\frac{1}{2}\sum_i \left\{
  \log2\pi(A_1^2 \lambda_i + \sigma^2) + \frac{c_i^2}{A_1^2\lambda_i +
  \sigma^2} \right\}.
\end{equation}
This formulation assumes both the GW signal and the white noise follow
Gaussian statistics; in this case each component value $c_i$ is drawn
from a normal distribution with zero mean and variance equal to
$A_1^2\lambda_i + \sigma^2$.  The different components are by
construction independent due to the choice of basis.  The product of all
the component distributions for a given pulsar results in the likelihood
function of Eqn.~\ref{eqn:logl}.  Using uniform/positive priors on
$\sigma$ and $A_1$, and marginalizing over $\sigma$, we convert the
likelihood function above into a posterior probability distribution for
$A_1$, shown in Figure~\ref{fig:amp_pdf}.  For the data from
J1713$+$0747, 95\% of the distribution falls under
$A_1<1.1\times10^{-14}$ for an $\alpha=-2/3$ spectrum.  This is the most
constraining single pulsar in our set.  Similar upper limits computed
for each of the pulsars at a variety of astrophysically relevant
$\alpha$ are listed in Table~\ref{tab:apdf}.

While we have framed the discussion so far in terms of a GWB signal,
this analysis method is more fundamentally a way to quantify timing
perturbations of a known/assumed spectral shape that may exist in the
data, regardless of their cause.  In this context we can look at these
single-pulsar results as a test for ``red'' power-law timing noise in
the data set.  In addition to the 95\% upper limit on $A_1$ already
described, two other statistics are useful:  The maximum-likelihood
estimate $\hat{A_1}$ and the ratio
$R=L(0,\hat{\sigma_0})/L(\hat{A_1},\hat{\sigma})$ between the maximum
likelihood values obtained by either fixing $A_1=0$ or allowing it to
vary.  $\hat{A_1}$ characterizes the strength of the red noise, and $R$
characterizes its significance in the data.  Values of $R$ near 1
indicate consistency with a white-noise-only model, while very small
values of $R$ indicate significant red noise is detected.  These three
values are listed for all pulsars for a selection of various expected GW
spectral indices in Table~\ref{tab:apdf}.  We see that two of the
pulsars -- J1643$-$1224 and J1910$+$1256 -- show very significant red
noise ($R<0.01$), and two others -- J1640$+$2224 and B1953$+$29 -- show
mildly significant red noise ($0.01\lesssim R < 0.1$).  The remaining 13
sources are consistent with white noise only.  Additional detailed
analyses of the noise in these data are ongoing, and include both a
Bayesian analysis covering a wider range of power-law spectral index
\citep{ellis:cov}, and an application of the \citet{coles:gls}
methodology to our data set \citep{perrodin:noise}.

We will not attempt to definitively determine the cause of the red noise
in the four sources noted here.  However we speculate that at least for
three of them a likely source is the interstellar medium, rather than
processes intrinsic to the pulsars.  As explained above, J1910$+$1256 and
B1953$+$29 do not include DM($t$) fits so their timing will be affected
at some level by unmodelled DM variation.  J1643$-$1224 has the second
highest DM of any pulsar in our set (after B1953$+$29), and has a high
predicted level of interstellar scatter-broadening, suggesting multipath
ISM effects may be an issue here.

\begin{table*}[tp]
\caption{\label{tab:apdf} Results from single-pulsar GW analyses.}
\begin{center}
\begin{tabular}{c|rrr|rrr|rrr}
\hline
Source & 
  \multicolumn{3}{|c}{$\alpha=-2/3$} & 
  \multicolumn{3}{|c}{$\alpha=-1$} & 
  \multicolumn{3}{|c}{$\alpha=-7/6$} \\
       & $\log_{10} R$ & Max-L $\hat{A_1}$ & 95\% $A_1$ limit 
       & $\log_{10} R$ & Max-L $\hat{A_1}$ & 95\% $A_1$ limit 
       & $\log_{10} R$ & Max-L $\hat{A_1}$ & 95\% $A_1$ limit \\
       &          &  ($10^{-15}$) & ($10^{-15}$) 
       &          &  ($10^{-15}$) & ($10^{-15}$) 
       &          &  ($10^{-15}$) & ($10^{-15}$) \\
\hline
J0030+0451 & $0.00$ & $0.0$ & $76.5$ & $0.00$ & $0.0$ & $47.8$ & $0.00$ & $0.0$ & $37.8$ \\
J0613--0200 & $0.00$ & $0.0$ & $100.9$ & $0.00$ & $0.0$ & $63.1$ & $0.00$ & $0.0$ & $49.9$ \\
J1012+5307 & $0.00$ & $0.0$ & $153.8$ & $0.00$ & $0.0$ & $109.7$ & $0.00$ & $0.0$ & $93.5$ \\
J1455--3330 & $0.00$ & $0.0$ & $242.8$ & $0.00$ & $0.0$ & $162.6$ & $0.00$ & $0.0$ & $133.1$ \\
J1600--3053 & $-0.00$ & $58.9$ & $973.9$ & $-0.01$ & $62.6$ & $1173.6$ & $-0.03$ & $68.1$ & $1240.4$ \\
J1640+2224 & $-1.31$ & $68.1$ & $501.6$ & $-1.84$ & $40.3$ & $413.3$ & $-1.90$ & $28.1$ & $350.2$ \\
J1643--1224 & $-4.57$ & $182.9$ & $455.3$ & $-4.57$ & $122.5$ & $333.6$ & $-4.70$ & $99.5$ & $284.6$ \\
J1713+0747 & $-0.22$ & $1.9$ & $11.3$ & $-0.22$ & $1.1$ & $7.6$ & $-0.29$ & $0.8$ & $6.2$ \\
J1744--1134 & $0.00$ & $0.0$ & $184.1$ & $0.00$ & $0.0$ & $119.1$ & $0.00$ & $0.0$ & $96.2$ \\
J1853+1308 & $-0.45$ & $23.6$ & $131.2$ & $-0.36$ & $8.3$ & $86.7$ & $-0.35$ & $5.3$ & $70.0$ \\
B1855+09 & $-0.31$ & $14.0$ & $70.0$ & $-0.59$ & $8.6$ & $45.6$ & $-0.65$ & $6.4$ & $36.3$ \\
J1909--3744 & $0.00$ & $0.0$ & $39.4$ & $0.00$ & $0.0$ & $26.8$ & $-0.00$ & $1.0$ & $25.0$ \\
J1910+1256 & $-4.66$ & $41.7$ & $229.7$ & $-4.91$ & $23.6$ & $144.6$ & $-4.85$ & $17.4$ & $115.1$ \\
J1918--0642 & $-0.31$ & $64.8$ & $545.0$ & $-0.21$ & $38.9$ & $383.1$ & $-0.11$ & $10.5$ & $282.6$ \\
B1953+29 & $-2.11$ & $274.9$ & $1877.7$ & $-1.43$ & $188.0$ & $1249.0$ & $-1.52$ & $160.4$ & $1110.5$ \\
J2145--0750 & $0.00$ & $0.0$ & $568.0$ & $0.00$ & $0.0$ & $372.6$ & $0.00$ & $0.0$ & $296.6$ \\
J2317+1439 & $-0.20$ & $43.4$ & $383.1$ & $-0.13$ & $26.4$ & $195.9$ & $-0.07$ & $20.9$ & $148.6$ \\
\hline
B1855$+$09 (KTR94)    & $-0.27$ & $5.6$ & $31.3$ & $-0.15$ & $1.8$ & $19.3$ & $-0.12$ & $1.1$ & $15.2$ \\
B1855$+$09 (combined) & $-8.53$ & $5.0$ & $13.1$ & $-7.09$ & $1.2$ &  $6.5$ & $-7.41$ & $0.7$ & $4.8$ \\
\hline
\end{tabular}

\end{center}
\end{table*}

\subsection{Verification}
\label{sec:gw:verify}

\begin{figure*}[tp]
\begin{center}
\plotone{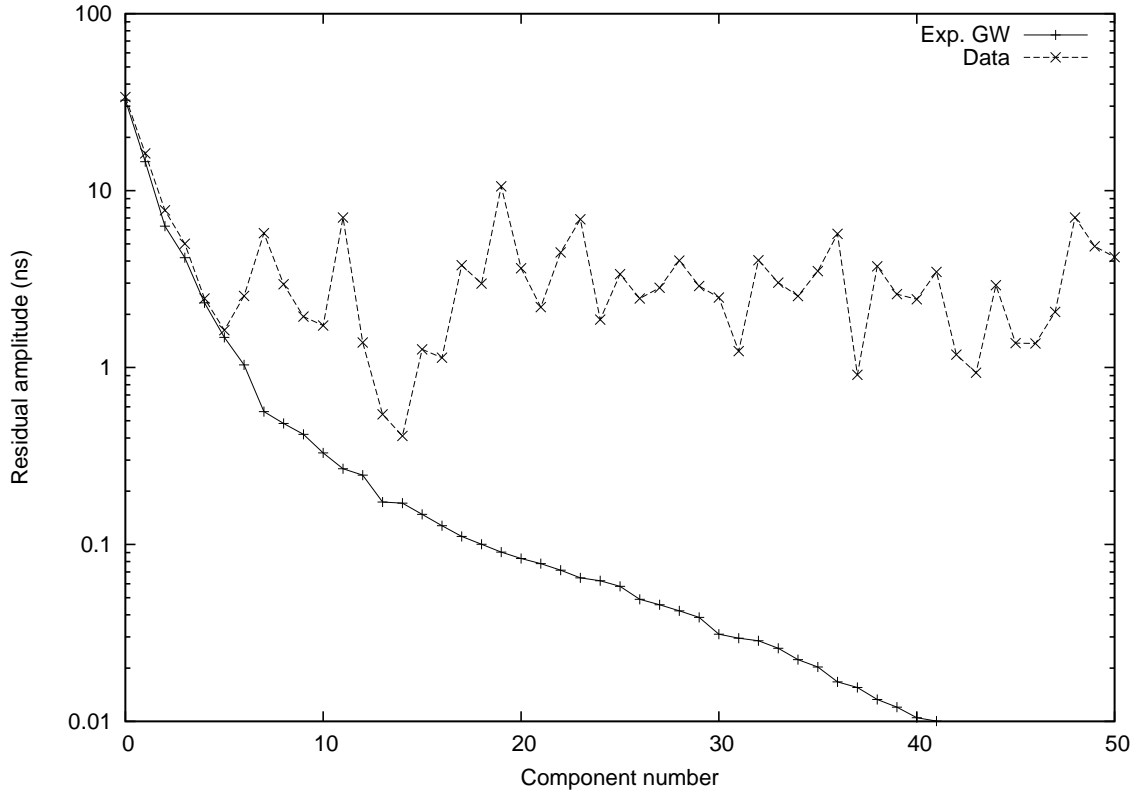}
\end{center}
\caption{\label{fig:sim_evals}
Eigenvalue spectra for J1713$+$0747 with simulated GW injected at an
amplitude of $A_1=10^{-14}$.  As in Figure~\ref{fig:evals}, the ``Exp.
GW'' line shows the expected GW amplitude determined from the eigenvalue
analysis, scaled to the known signal level.  The ``Data'' line shows the
root-mean-square residual in each component, averaged over 1000
realizations of the simulation.  The correct GW signal level is observed
in the data, until the level drops below the white noise near component
7.
}
\end{figure*}

\begin{figure*}[tp]
\begin{center}
\plotone{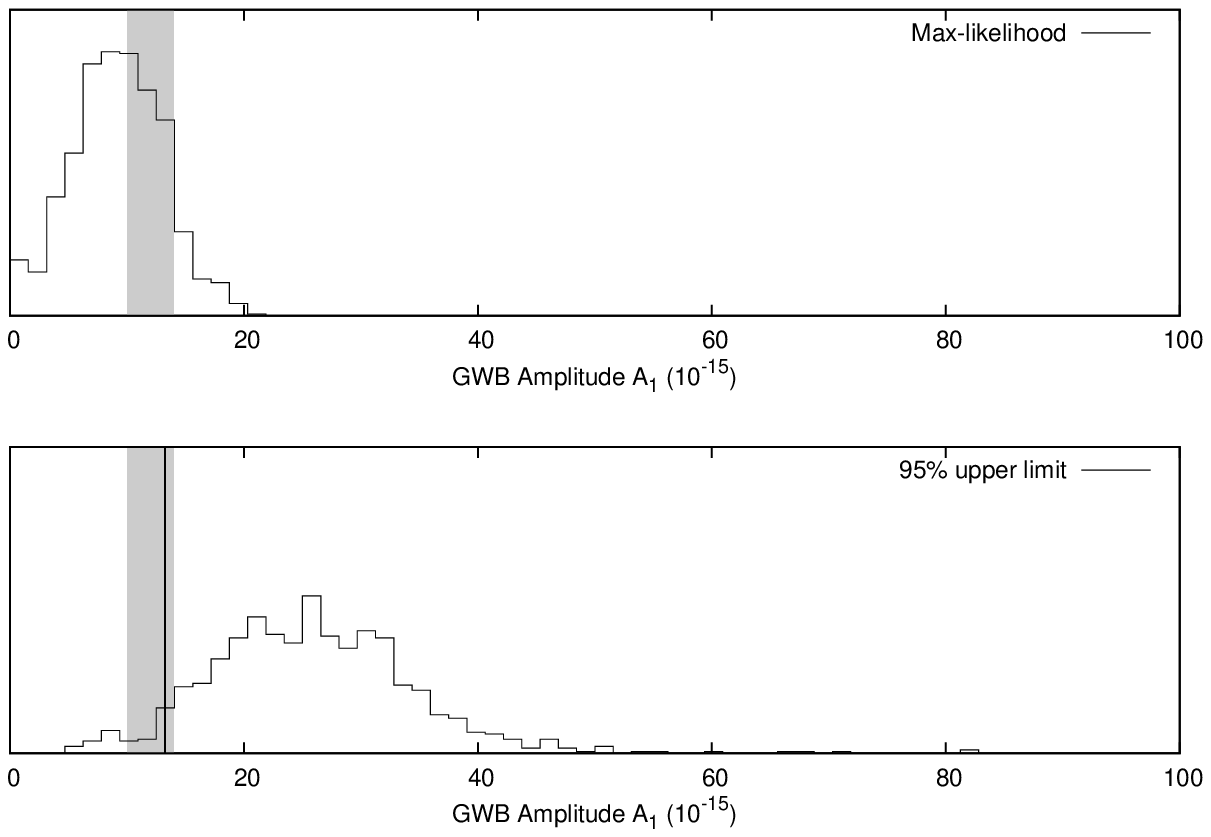}
\end{center}
\caption{\label{fig:limit_hist}
Histograms of maximum-likelihood GW amplitude (upper panel), and 95\%
upper limit (lower panel) computed from 1000 realizations of a simulated
$A_1=10^{-14}$, $\alpha=-2/3$ GWB added to J1713$+$0747 timing data.
The shaded range represents the true GW signal level, including both the
simulated signal and any existing GW signal in the data, up to the limit
presented in \S\ref{sec:gw:single}.  The vertical line in the lower
panel shows the 5\% quantile of the histogram values.
}
\end{figure*}

The methods discussed in the previous section do not rely at all on
simulated data to produce a GW upper limit or red noise estimate.
However, it is still a valuable exercise to analyze a simulated GW
signal using this approach, as a test of its validity.  To do so, we
added simulated GW to our timing data for J1713$+$0747, using the {\sc
TEMPO2 GWbkgrd} plugin \citep{hobbs:tempo2_gw}.  GW were added using
$\alpha=-2/3$ and $A_1=10^{-14}$.  These data were then analyzed using
the eigenvalue procedure described above.  Figure~\ref{fig:sim_evals}
shows the root-mean-square component values ($c_i$) from 1000 GW
realizations, along with the square root of the associated eigenvalues
($\lambda_i^{1/2}$) scaled to the known GW signal level.  The excellent
agreement between these two lines shows that our algorithm is correctly
predicting the amount of GW power present in the timing residuals.
Figure~\ref{fig:limit_hist} presents results of computing
maximum-likelihood and 95\% upper limit amplitude values from the
simulated data.  These statistics agree well with the known amount of GW
signal in the simulated data (grey region in plots).

As an additional test of this method, we analyzed the publicly-available
seven-year B1855$+$09 timing data set published by \citet[][hereafter
KTR94]{kaspi:1937}, from observations taken during 1986--1992.  This
data set has been used for several previous GW analyses, so provides a
useful benchmark for testing different analysis methods.  For this
comparison we focus on the GW spectrum originally considered by KTR94,
with $\alpha=-1$.  From the KTR94 data set alone we find no significant
evidence for red timing noise (see results in Table~\ref{tab:apdf}), and
using our method it sets a 95\% upper limit of $A_1<1.9\times10^{-14}$.
This is somewhat more conservative than the published limits of
$A_1<1.3\times10^{-14}$ (KTR94) and $A_1<1.4\times10^{-14}$
\citep{jenet:limit} obtained from the same data.\footnote{Limits
originally published in units of $\Omega_{gw} h^2$ were converted to
equivalent $A_1$ using Eqn.~3 of \citet{jenet:limit}.} This difference
may be due to the varying strategies used by these three analyses to
account for effects of the timing model fit and the possibility of
existing red noise in the data.

\begin{figure*}[p]
    \begin{center}
    \plotone{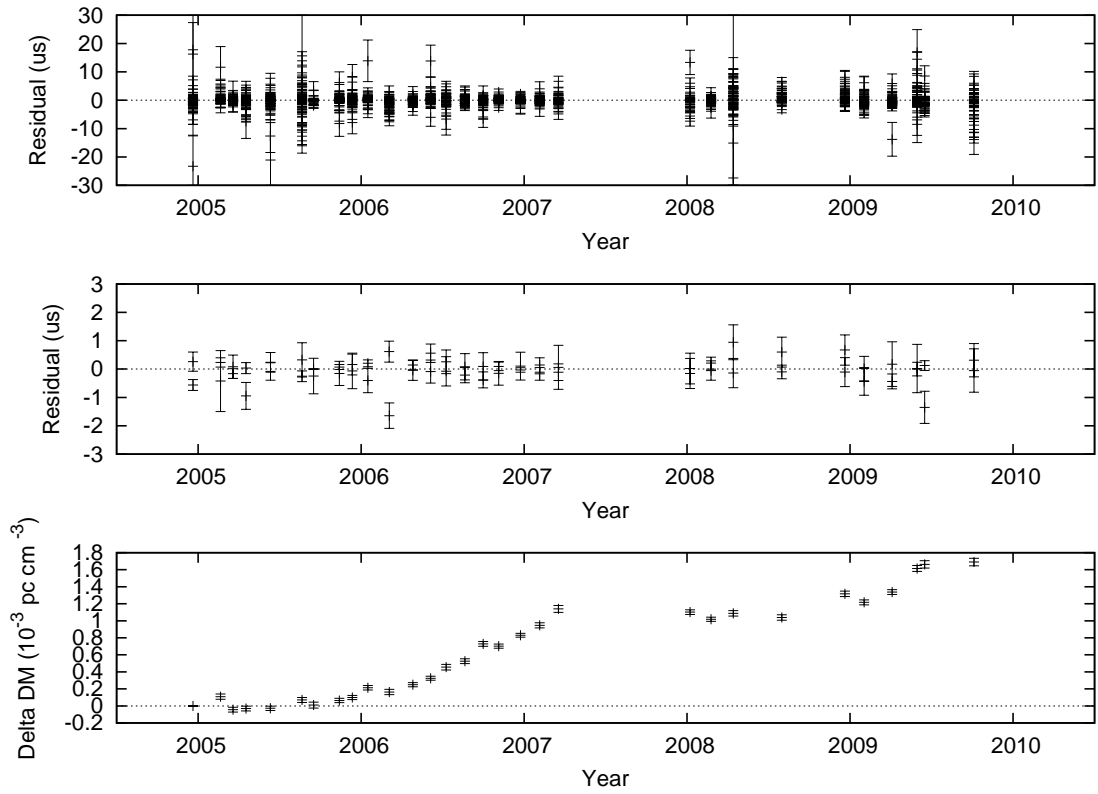}
    \caption{\label{fig:1855:timing}
    Timing summary for PSR~B1855$+$09, see Figure~\ref{fig:1713:timing} for details.
    }
    \end{center}
    \end{figure*}

Combining the KTR94 data with our newer observations of this pulsar into
a single analysis produces a limit $A_1<6.5\times10^{-15}$, comparable
to what we get from our J1713$+$0747 data alone.  However, very
significant red noise is detected in the combined B1855$+$09 dataset
(Table~\ref{tab:apdf}).  A fundamental feature of red noise processes is
that their amplitude grows with increasing data span.  Therefore this
detection in the combined data set is not inconsistent with the lack of
such a detection in either of the two data subsets on their own.
Another contributing factor may be unmodeled variation in DM.  The KTR94
data were obtained at a single frequency and as such cannot be corrected
for DM variation.  In our newer data, a significant DM$(t)$ trend is
clearly visible for this pulsar (Figure~\ref{fig:1855:timing}).

\subsection{Cross-correlation Analysis}
\label{sec:gw:xcorr}

To detect a GW signal present in the timing data, we must consider
correlations among the timing residuals of different pulsars, as
previously discussed.  This is complicated by the fact that residuals
from each pulsar are produced by separate timing model fits.  The fits
potentially incorporate a wide variety of effects (e.g. binary
parameters) that vary from pulsar to pulsar.  Furthermore the span and
cadence of observations is not necessarily identical between all the
sources, and each may be affected by a different level of red timing
noise from either GW or other sources.  Correctly measuring the
correlation between two such residual timeseries is therefore more
complicated than simply multiplying and averaging the two signals, as
would be done in a standard correlation analysis.

Referring to the discussion in \S\ref{sec:gw:timing} and in particular
Eqn.~\ref{eqn:r_xcov}, given the actual timing model fits used for any
two pulsars and an expected gravitational wave spectrum, we determine
the expected cross-correlation for the GW signal between all pairs of
residuals of each pulsar pair ($\m{C_{r_a,r_b}^{GW}}$).  We use the
elements of this matrix, along with the estimated noise parameters from
\S\ref{sec:gw:single} as weights for a cross-correlation sum to estimate
the GW signal power correlated between pulsars $a$ and $b$:
\begin{equation}
\label{eqn:xcorr}
  \rho_{ab} = \frac{\sum_{ijkl} 
    r_i^{(a)} 
    \left(C^{tot(a)}\right)^{-1}_{ij}
    C_{jk}^{GW (a,b)}
    \left(C^{tot(b)}\right)^{-1}_{kl}
    r_l^{(b)}} 
    {\sum_{ijkl}
    \left(C^{tot(a)}\right)^{-1}_{ij}
    C_{jk}^{GW (a,b)}
    \left(C^{tot(b)}\right)^{-1}_{kl}
    C_{il}^{GW (a,b)}
    }.
\end{equation}
The components of this expression are:
\begin{itemize}

  \item The sum indices $i$ and $j$ run from 1 to $N_a$, the number of
  TOAs for pulsar $a$.  Similarly, $k$ and $l$ run from 1 to $N_b$.

  \item $r_i^{(a)}$ and $r_l^{(b)}$ are simply the post-fit timing
  residuals for pulsars $a$ and $b$ respectively.

  \item $C_{ij}^{tot(a)}$ and $C_{kl}^{tot(b)}$ are the {\it total}
  post-fit covariance matrices for each pulsar.  These are determined
  using the maximum-likelihood red noise parameters found in the earlier
  single-pulsar analysis, combined with the TOA measurement
  uncertainties, and modified by the timing model fit (see
  Eqn.~\ref{eqn:r_cov}):
  \begin{equation}
    \m{C^{tot}_r} = \m{R} \left( \hat{A_1}^2 \m{C^{GW}_y} + \m{W}^{-1} \right)
    \m{R}^T
  \end{equation}

  \item $C_{jk}^{GW(a,b)}$ are the elements of $\m{C_{r_a,r_b}^{GW}}$,
  the expected post-fit cross-covariance matrix for the GW signal in
  pulsars $a$ and $b$, computed using Eqn.~\ref{eqn:r_xcov} and the
  assumed power-law spectral shape.

\end{itemize}
The presence of the inverse covariance matrices in Eqn.~\ref{eqn:xcorr}
effectively serves to ``whiten'' the timing residuals before computing
the cross-correlation sum.  This reduces the weight of data from pulsars
that have red noise components, potentially including any ``self-noise''
contribution from the stochastic GWB itself.  This formulation is
conceptually very similar to the frequency-domain optimal statistic
presented by \citet{anholm:opt}.  

The post-fit covariance matrices $\m{C^{tot}_r}$ are singular, due to
the degrees of freedom removed by the timing fit, or equivalently by the
application of the $\m{R}$ operator.  Therefore the matrix inverse
technically does not exist.  In this situation it is appropriate to use
a singular value decomposition-based pseudoinverse in its place.  This
is equivalent to performing the correlation in the subspace of signals
that are orthogonal to (i.e., not absorbed by) the timing fit.  The
uncertainty on $\rho_{ab}$ is given by:
\begin{equation}
\label{eqn:xcorr_err}
  \sigma_{\rho_{ab}} = 
    \left(
    \sum_{ijkl}
    \left(C^{tot(a)}\right)^{-1}_{ij}
    C_{jk}^{GW (a,b)}
    \left(C^{tot(b)}\right)^{-1}_{kl}
    C_{il}^{GW (a,b)}
    \right)^{-1/2}.
\end{equation}

\begin{figure*}[htp]
\begin{center}
\plotone{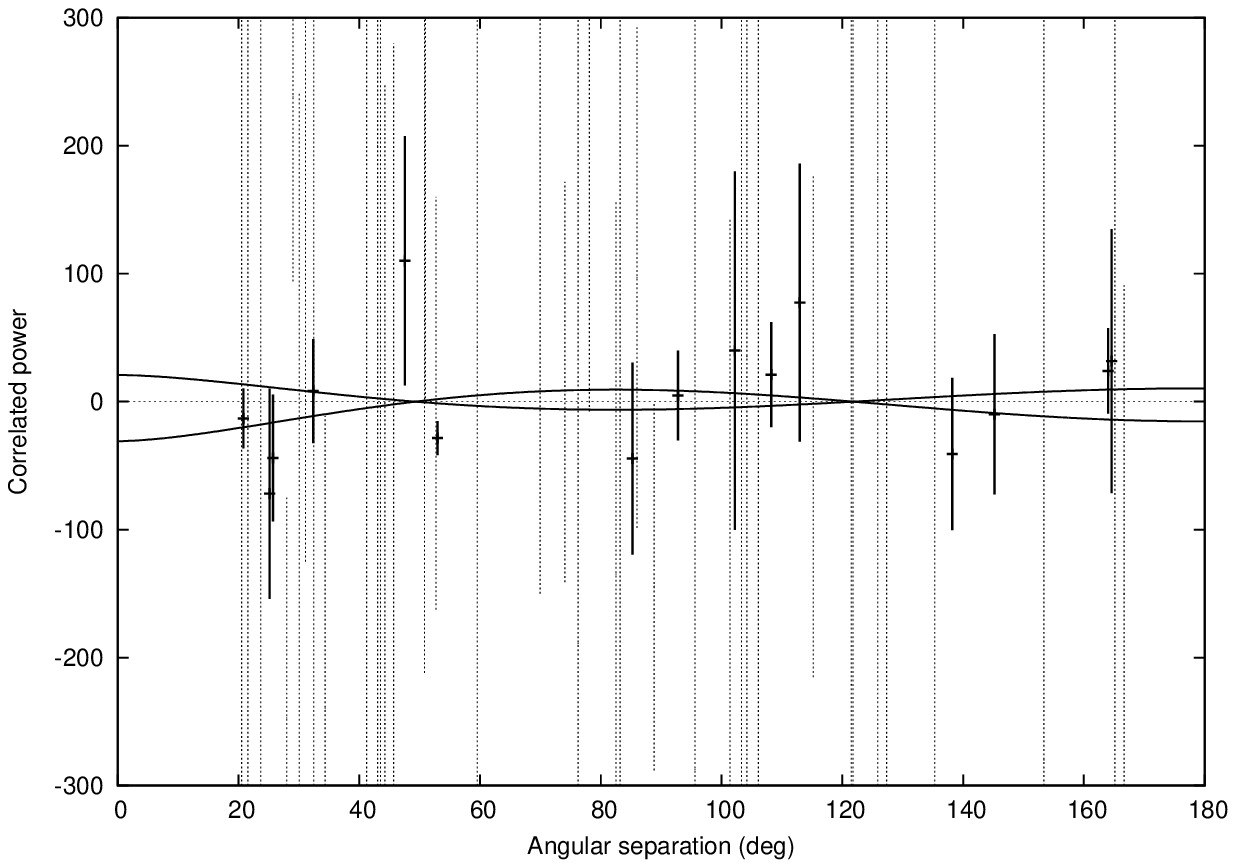}
\end{center}
\caption{\label{fig:xcorr}
Measured cross-correlated power $\rho_{ab}$ as a function of separation
angle $\theta_{ab}$ for pairs of pulsars in our set, with error bars
showing 1-$\sigma$ uncertainty.  Power is normalized relative to an
$A_1^2=10^{-30}$, $\alpha=-2/3$ GWB.  The lines show the $\pm$2~$\sigma$
fit to the amplitude of the Hellings-Downs function $\zeta(\theta)$.
All 136 cross-correlation points were used for the fit, however for
clarity the 15 lowest-uncertainty values are denoted with solid/bold
symbols.  
}
\end{figure*}

The resulting $\rho_{ab}$ for all 136 pulsar pairs, with $\alpha=-2/3$,
is shown in Fig.~\ref{fig:xcorr}.  We search for the presence of the GWB
in these data by fitting them to an amplitude (proportional to $A_1^2$)
times the Hellings-Downs angular function $\zeta(\theta_{ab})$.  The
best fit, also shown in Fig.~\ref{fig:xcorr}, results in $A_1^2 = (-10
\pm 26)\times10^{-30}$ for $\alpha=-2/3$.  This is consistent with no
detectable GWB in the data, and also can be interpreted as a 2-$\sigma$
upper limit of $A_1 < 7.2\times10^{-15}$, an improvement over the
single-pulsar limits presented in \S\ref{sec:gw:single}.  The
reduced-$\chi^2$ of this fit is 0.95, providing additional confidence
that the uncertainty estimate of Eqn.~\ref{eqn:xcorr_err} is correct.

A table of the top 15 correlation measurements, sorted by increasing
uncertainty, is presented in Table~\ref{tab:xcorr}.  It is clear that
the measurement is dominated by pairs involving one or both of the two
best-timing pulsars in the set, J1713$+$0747 and J1909$-$3744.  In
total, 8 separate pulsars contribute to these 15 lowest-uncertainty
points.  The limit being set primarily by a small number of pulsars is
due to the fact that our current data set is in a white noise dominated,
low GW amplitude regime.  As the length and quality of such datasets
increase, they are expected to become increasingly dominated by red
noise components, either from the GWB itself or from intrinsic pulsar
timing noise.  In the red noise dominated and/or strong GW regime, a
large number of pulsars (likely $\sim$20-50) is required to improve the
significance of a GW detection \citep{jenet:gw, cordes:gw_detect}.  This
provides motivation for continuing to observe many MSPs, even if upper
limits are currently only dominated by a small subset of the pulsars.

\begin{figure}[htp]
\begin{center}
\plotone{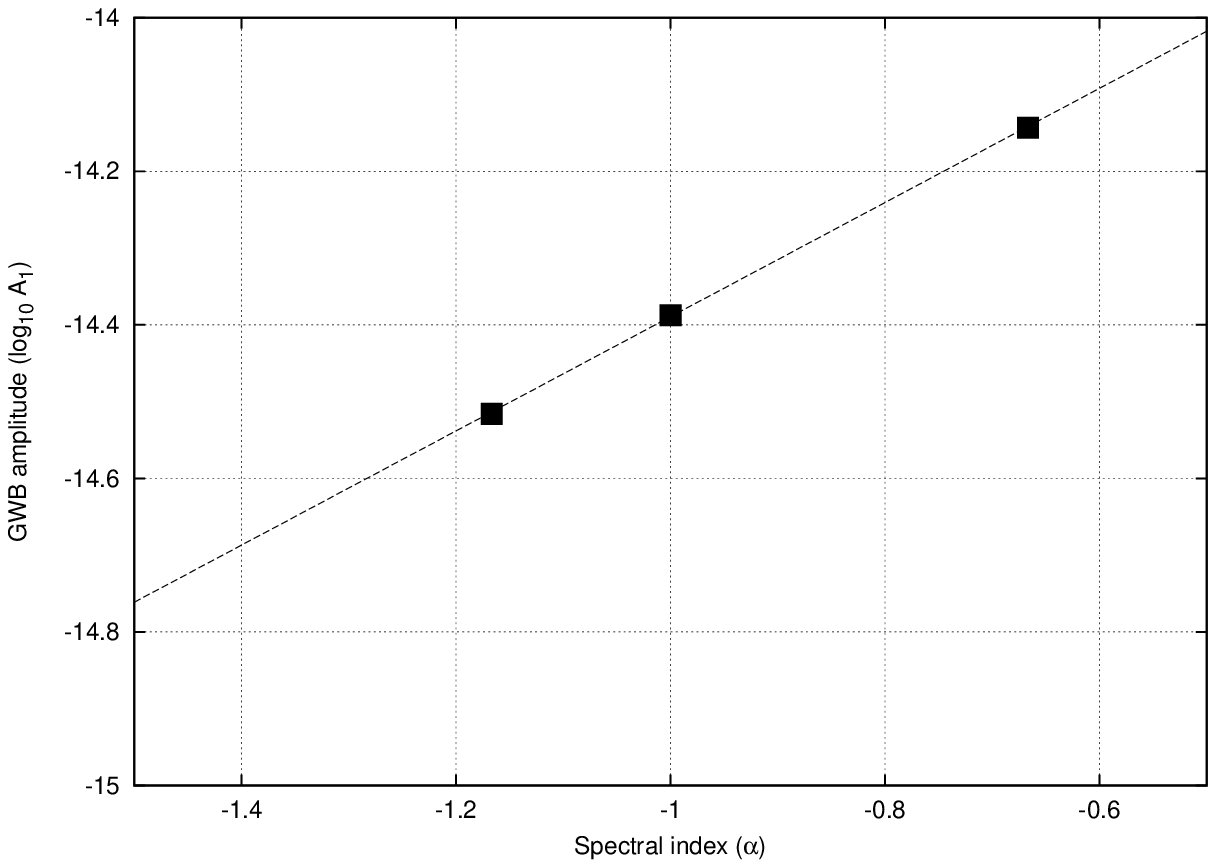}
\end{center}
\caption{\label{fig:alpha_scale}
Measured 2-$\sigma$ upper limits on $A_1$ as a function of GW spectral
index $\alpha$ (squares).  The values are consistent with a simple
scaling based on $A_T=2.26\times10^{-14}$ at $T=5.54$~years and
Eqn.~\ref{eqn:alpha_scale} (line).  This relationship can be used to
convert these results to equivalent limits for other values of $\alpha$.
}
\end{figure}

Table~\ref{tab:xcorr} also lists the best-fit $A_1^2$ and its
uncertainty for three different spectral indices.  The resulting
2-$\sigma$ upper limits are $A_1= 7.2\times10^{-15}$,
$4.1\times10^{-15}$, and $3.0\times10^{-15}$ for $\alpha=-2/3$, $-1$ and
$-7/6$ respectively.  The change in value of $A_1$ with $\alpha$ is
primarily due to the fact that although we are using 1~year$^{-1}$ as
the reference GW frequency, the measurement is most sensitive to
frequencies near $T^{-1}$, where $T\sim5$~years is the effective length
of the multi-pulsar data set.  Therefore, we expect the various limits
on $A_1$ to scale with $\alpha$ as follows:
\begin{equation}
\label{eqn:alpha_scale}
  A_1(\alpha) = A_T \left( \frac{T}{1~\mathrm{year}} \right)^\alpha 
\end{equation}
Using our three measured $A_1$ limits to empirically determine $A_T$ and
$T$ gives $A_T=2.26\times10^{-14}$ and $T=5.54$~years (see
Figure~\ref{fig:alpha_scale}).  These values can be used togther with
Eqn.~\ref{eqn:alpha_scale} to convert our measurement into an
approximate limit on $A_1$ for values of $\alpha$ that we have not
considered here.

\begin{table*}[tp]
\caption{\label{tab:xcorr} Cross-correlated power measurements and GW
results.}
\begin{center}
\begin{tabular}{cc|r|rr|rr|rr}
\hline
Pulsar $a$ & Pulsar $b$ & Angle $\theta_{ab}$ & 
  \multicolumn{2}{|c}{$\alpha=-2/3$} &
  \multicolumn{2}{|c}{$\alpha=-1$} &
  \multicolumn{2}{|c}{$\alpha=-7/6$} \\
           &            &                     & 
                                                Cross-power & Uncertainty &
                                                Cross-power & Uncertainty &
                                                Cross-power & Uncertainty \\
  &  & (deg) & $\rho_{ab}$ (10$^{-30}$) & $\sigma_{\rho_{ab}}$ (10$^{-30}$) 
             & $\rho_{ab}$ (10$^{-30}$) & $\sigma_{\rho_{ab}}$ (10$^{-30}$) 
             & $\rho_{ab}$ (10$^{-30}$) & $\sigma_{\rho_{ab}}$ (10$^{-30}$) \\
\hline
 J1713$+$0747 & J1909$-$3744 & 53.0 & -28.4 & 13.2 & -8.8 & 4.4 & -5.9 & 2.7 \\
 J1713$+$0747 & J1744$-$1134 & 20.8 & -13.1 & 23.4 & -3.5 & 7.7 & -1.6 & 4.0 \\
 J0613$-$0200 & J1713$+$0747 & 164.0 & 24.0 & 33.4 & 7.1 & 11.0 & 3.8 & 5.9 \\
 J1012$+$5307 & J1713$+$0747 & 92.8 & 4.8 & 35.0 & 1.1 & 11.2 & 0.6 & 6.0 \\
 J1744$-$1134 & J1909$-$3744 & 32.4 & 8.3 & 40.7 & -0.6 & 12.6 & 0.4 & 8.0 \\
 J0030$+$0451 & J1713$+$0747 & 108.2 & 21.2 & 41.0 & 6.7 & 12.6 & 3.8 & 6.5 \\
 J1713$+$0747 & B1855$+$09 & 25.7 & -44.0 & 49.5 & -11.5 & 17.9 & -5.5 & 9.9 \\
 J0613$-$0200 & J1909$-$3744 & 138.2 & -40.9 & 59.4 & -13.0 & 18.3 & -7.8 & 11.7 \\
 J1012$+$5307 & J1909$-$3744 & 145.2 & -9.9 & 62.6 & -4.8 & 18.7 & -3.1 & 11.9 \\
 J0030$+$0451 & J1909$-$3744 & 85.3 & -44.4 & 75.0 & -14.4 & 21.2 & -8.3 & 13.0 \\
 J1713$+$0747 & J1853$+$1308 & 25.2 & -71.8 & 82.1 & -18.7 & 19.5 & -9.8 & 9.6 \\
 B1855$+$09 & J1909$-$3744 & 47.5 & 110.2 & 97.3 & 37.1 & 33.1 & 20.0 & 21.2 \\
 J0613$-$0200 & J1744$-$1134 & 164.6 & 31.6 & 103.2 & 9.3 & 31.3 & 5.2 & 16.8 \\
 J1012$+$5307 & J1744$-$1134 & 113.0 & 77.4 & 108.6 & 21.7 & 32.2 & 11.5 & 17.1 \\
 J0030$+$0451 & J1744$-$1134 & 102.2 & 39.9 & 140.0 & 11.6 & 38.2 & 6.4 & 19.4 \\
\multicolumn{2}{c|}{\ldots} & 
\ldots &
\multicolumn{2}{|c}{\ldots} & 
\multicolumn{2}{|c}{\ldots} & 
\multicolumn{2}{|c}{\ldots} \\
\hline
\multicolumn{3}{c|}{Best-fit $A_1^2$ (10$^{-30}$)} & 
  \multicolumn{2}{|c}{$-10 \pm 26$} &
  \multicolumn{2}{|c}{$-3.7 \pm 8.4$} &
  \multicolumn{2}{|c}{$-1.9 \pm 4.6$} \\
\hline
\end{tabular}

\end{center}
\end{table*}

\section{Discussion}
\label{sec:discuss}

\subsection{Astrophysical Consequences}
\label{sec:discuss:astro}

As mentioned previously, the strongest expected source of GW in the
nanohertz band is the ensemble background from supermassive black hole
binaries.  It is on these which we will focus our discussion here.  The
GWB predicted from the supermassive black hole binary population is
dependent on a number of observationally ill-constrained values: e.\,g.
the ubiquity of supermassive black holes in galaxy centers, the
prevalence of merging, as well as the timescale and dominant mode of
black hole growth (via accretion or merger with other black holes).  

Recently, \citet{sesana:gwb} considered a broad range of probable
parameter uncertainty values applied to four leading scenarios for black
hole growth in the context of hierarchical merging. Of most importance
here, they found that when accounting for uncertainties in all model
parameters, the estimated range of predicted $A_1$ values produced a
relatively narrow predicted amplitude range of $1\times10^{-16}\lesssim
A_1 \lesssim 3\times10^{-15}$.  While we are not yet sensitive enough to
impact even the upper boundary of their predicted range, it is pertinent
to note that our current sensitivity is approximately a factor of 2 away
from the expected highest prediction.  If adequate GW sensitivity can be
achieved, pulsar timing will probe astrophysical quantities not readily
accessible by electromagnetic observations.  The most uncertain factors
noted by \citet{sesana:gwb} were the supermassive back hole mass
function, the local merger rate of massive galaxies, the amount of
merger-induced accretion onto black holes, and the post-merger black
hole inspiral rate.  A measurement or constraint on the mass function
leads to limits on galactic host/black hole mass relationships,
particularly in the most massive regime.

The other relevant finding of \citet{sesana:gwb} was that with a paucity
depending on the input model, GW spectral bins above
$\sim$$1-5\times10^{-8}$\,Hz were populated by increasingly few black
holes -- at the highest frequencies, sometimes one or none.  A lack of
black holes contributing to the high-frequency GW spectrum could affect
us in two ways.   It could change the observed GWB spectral slope.  In
fact, \citet{sesana:gwb} uses an expanded version of our Eq.\,1 to
include a break in the spectrum.  Alternately it could induce
non-stochastic signals, requiring a directional/point-source analysis
approach to detect the emitted signal.  Either of these consequences
would corrupt the sensitivity and applicability of the analysis
performed here, \emph{particularly if the analysis were performed on
data spans of $\lesssim$3\,years}. All but one of the pulsar data sets
presented here exceed this length, thus the influence of this issue is
expected to be minimal.  As the contributing population is expected to
increase at lower frequencies, the impact of this problem will decrease
as further data are collected on these pulsars, and as we reach the
aforementioned sensitivity improvements that come with increased data
span.  Nonetheless, further studies must be performed to address the
effects of the $N$-source contribution issue, particularly to understand
its effect on the expected Hellings-Downs function.

\subsection{Cosmic Strings}
\label{sec:discuss:strings}

Cosmic strings or superstrings are linear-dimensional structures
proposed to arise due to phase transitions in the early universe.
Interactions between members of a network of such objects may create
cusps or loops that are precicted to radiate significant GW, producing a
stochastic GWB with $\alpha\simeq-7/6$ in the nanohertz frequency band
\citep[e.g.,][]{damour:string,siemens:string}.  The amount of GW
produced depends on physical parameters such as the string loop size
scale, the reconnection probability and the string tension.  Conversely,
measured GW limits constrain the possible values of these parameters.
Following the analysis of \citet{siemens:string}, our results for cosmic
strings take the form of sections of cosmic string parameter space that
are ruled out or allowed by our upper limit on the gravitational wave
background.

\begin{figure}[htp]
\begin{center}
\plotone{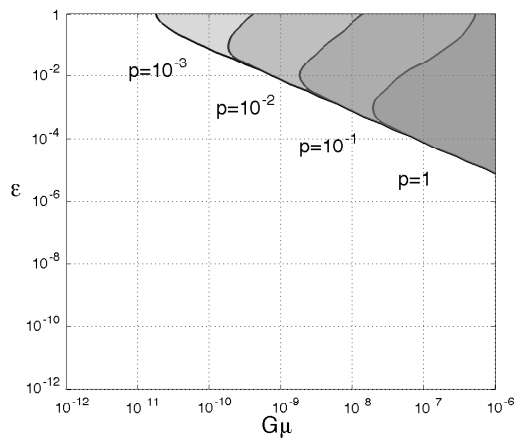}
\end{center}
\caption{\label{fig:small_loop} 
Cosmic string parameter space constraints from our measurement, in the 
small loop case.  The shaded areas shown regions of string tension
($G\mu$) and loop size ($\varepsilon$) that are ruled out by our
measurement for various values of reconnection probability ($p$).
}
\end{figure}

In the case where cosmic string loop sizes are set by gravitational back
reaction (``small loop'' case) we explore the three-dimensional
parameter space of cosmic string dimensionless tension $G\mu$,
reconnection probability $p$, and loop sizes parametrized by
$\varepsilon$ \citep[for details, see][]{siemens:string}. The shaded
areas of Fig.~\ref{fig:small_loop} show the regions of the
$\varepsilon$-$G\mu$ plane ruled out for four values of the reconnection
probability $p=1$, $10^{-1}$, $10^{-2}$ , and $10^{-3}$. For example,
for $\varepsilon=1$, and $p=10^{-3}$ the string tension $G\mu$ is
constrained to be less than about $2\times10^{-11}$.

\begin{figure}[htp]
\begin{center}
\plotone{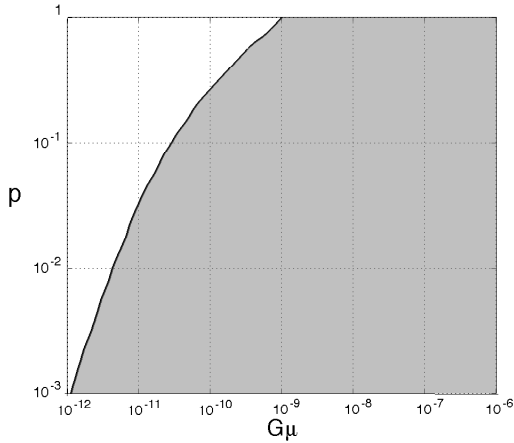}
\end{center}
\caption{\label{fig:large_loop}
Cosmic string constraints in the large loop case, in terms of string
tension $G\mu$ and reconnection probability $p$.  The shaded area is
ruled out by our GW upper limit.
}
\end{figure}

If the size of cosmic string loops is instead given by the large scale
dynamics of the network, as suggested by recent numerical
simulations~\citep{BlancoPillado:2011dq}, we fix the cosmic string loop
length and explore the two-dimensional parameter space of reconnection
probability and string tension following \citet{siemens:string}. These
results are shown in Fig~\ref{fig:large_loop}.  The stochastic
background produced in these models is substantially larger therefore
our constraints on parameters are correspondingly tighter. For example
for a reconnection probability $p=1$, $G\mu < 10^{-9}$, and for a
reconnection probability of $p=10^{-3}$ all cosmic string tensions above
$10^{-12}$ are ruled out.

\subsection{Future Performance}
\label{sec:discuss:future}

Predicting the future sensitivity of a pulsar timing array project to GW
depends on a large number of poorly constrained factors.  These include
the level of pulsar-intrinsic timing noise in the MSP population
\citep{shannon:msps}, the statistical characteristics of the pulsar
radio emission \citep{oslowski:snr}, and the influence of the
interstellar medium on pulsar timing \citep{coles:ism}.  A detailed
assessment of the effect of various kinds of noise on GW detection
sensitivty was done by \citet{cordes:gw_detect}.  Assuming all factors
besides experiment duration $T$ (e.g., number of pulsars, observing
cadence, telescope instrumentation, etc) remain the same, we can roughly
bound our future GW limits by two cases:
\begin{itemize}

  \item Most optimistically, the data could remain dominated primarily
  by ``white'' radiometer noise.  In this case, limits on $A_1$ scale in
  proportion to $T^{\alpha-3/2}$ For example, $A_1\propto T^{-13/6}$ for
  the supermassive black hole GWB.  With an additional five years of
  data, we will reach $A_1\sim1.6\times10^{-15}$ for this spectrum, well
  into the range of expected GW amplitudes.

  \item On the other hand, if the data are completely dominated by
  timing noise with spectral index equal to or steeper than the GW
  signal, limits improve only as $T^{-1/2}$.  Doubling the data span
  would only improve the result to $A_1\sim5\times10^{-15}$.  Note that
  this does not appear to be the situation currently for the majority of
  our pulsars.

\end{itemize} 
Intermediate cases, which are more likely to apply in reality than
either of the two extremes above, include those where only some of the
pulsars are timing-noise dominated, or have timing noise that is less
red than the GW signal.  Note that these scalings only apply for GW {\it
limits} (that is, non-detections) -- in the GW signal-dominated regime,
and during the transition from limit to detection, different rules apply
\citep[see][]{cordes:gw_detect}.  These estimates also do not take into
account any future improvements in the experiment.  In reality both
telescopes have recently undergone a major upgrade in digital pulsar
instrumentation with the installation of the GUPPI and PUPPI pulsar
backends \citep{demorest:guppi}.  These provide an order of magnitude
more radio bandwidth than was used for this analysis.  NANOGrav is also
currently in the process of expanding the number of pulsars monitored to
$\sim$30, as new MSPs are discovered by searches including {\it Fermi}
gamma-ray source follow-up \citep[e.g.,][]{ransom:fermi}, the Arecibo
PALFA survey \citep{kaspi:palfa}, and the Green Bank North Celestial Cap
survey \citep{stovall:gbncc}.  Additional improvements may come from
connecting our current data set to historical pulsar timing data
stretching back as far as 20 years for some sources.\\

\section{Conclusions}
\label{sec:conc}

In this paper, we presented and analyzed five years of radio timing data
on 17 pulsars taken with the two largest single-dish telescopes in the
world.  Our timing analysis included novel methods for dealing with
time-variable dispersion measure and intrinsic profile shape evolution
with frequency.  We achieved sub-microsecond RMS timing results on all
but two of our sources, and RMS residuals of only $\sim$30--50~ns in the
two best cases.  We presented a new time-domain method for detecting
and/or limiting timing flucutations of a known spectral shape but
unknown amplitude.  The key feature of this analysis was proper
accounting for the signal power removed by the timing model fit, without
requiring dependence on simulation for this step.  Applied to our timing
of J1713$+$0747, these methods set a single-pulsar limit on the
stochastic gravitational wave background amplitude of
$A_1<1.1\times10^{-14}$ (95\%) for the expected supermassive black hole
GWB spectrum with spectral index $\alpha=-2/3$.  We discussed how to
measure cross-correlations between the timing of pairs of pulsars while
also accounting for both the timing model fit and the presence of
correlated (non-white) noise in the data.  For $\alpha=-2/3$, the
measured cross-correlations in our data set constrain $A_1^2$ to
$(-10\pm26)\times10^{-30}$, or alternately a 2-$\sigma$ upper limit of
$A_1<7.2\times10^{-15}$.  We discussed how our measurement will improve
with time, and suggest that prospects are good for obtaining
astrophyiscally constraining GW limits, or possibly even a detection,
over the next five years.

\acknowledgements The NANOGrav project receives support from the
National Science Foundation (NSF) PIRE program award number 0968296.
NANOGrav research at UBC is supported by an NSERC Discovery Grant and
Discovery Accelerator Supplement.  P.B.D. acknowledges support from a
Jansky Fellowship of the National Radio Astronomy Observatory during
2007--2010.  A.N.L. gratefully acknowledges the support of NSF grant AST
CAREER 07-48580.  Part of this research was carried out at the Jet
Propulsion Laboratory, California Institute of Technology, under a
contract with the National Aeronautics and Space Administration.  The
National Radio Astronomy Observatory is a facility of the NSF operated
under cooperative agreement by Associated Universities, Inc.  The
Arecibo Observatory is operated by SRI International under a cooperative
agreement with the NSF (AST-1100968), and in alliance with Ana G.
M\'{e}ndez-Universidad Metropolitana, and the Universities Space
Research Association.\\

\bibliographystyle{apj}
\bibliography{gw}

    \begin{figure*}[p]
    \begin{center}
    \plotone{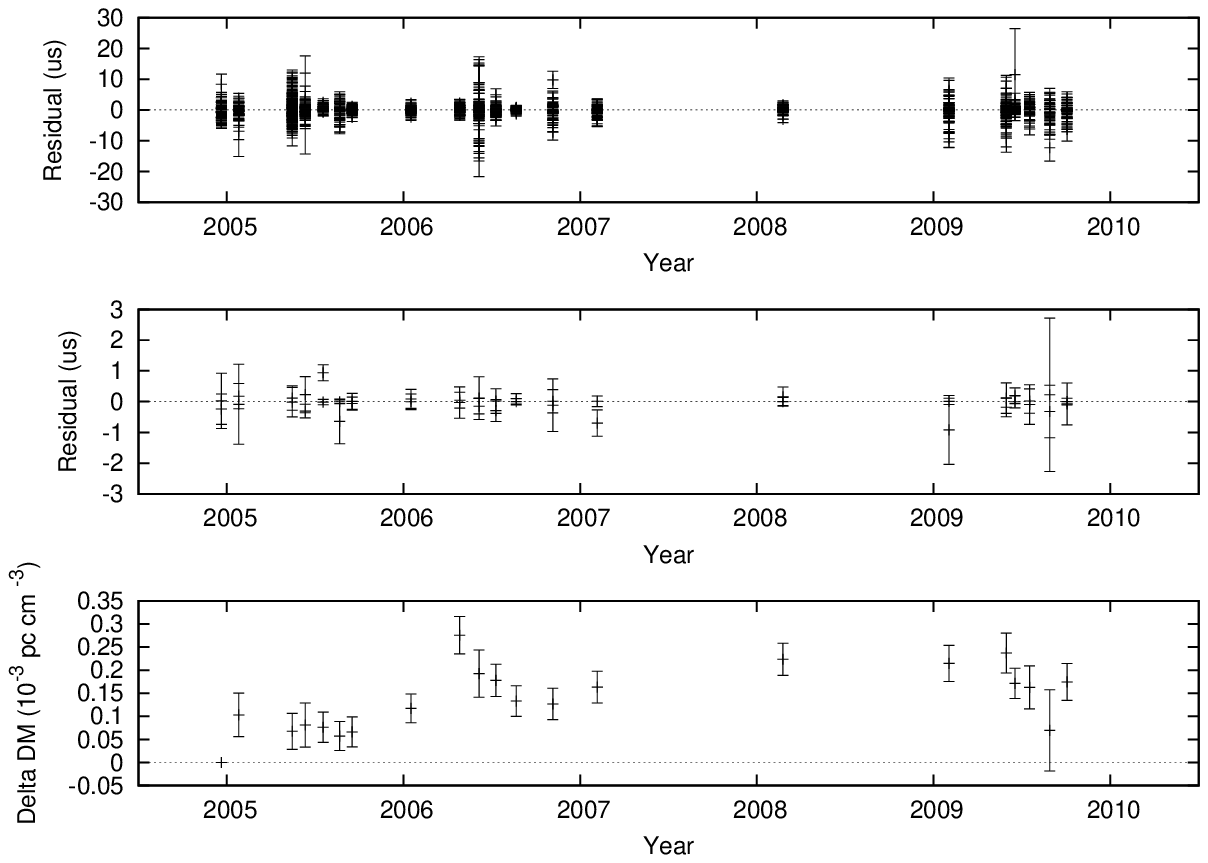}
    \caption{\label{fig:0030:timing}
    Timing summary for PSR~J0030$+$0451, see Figure~\ref{fig:1713:timing} for details.
    }
    \end{center}
    \end{figure*}

    \begin{figure*}[p]
    \begin{center}
    \plotone{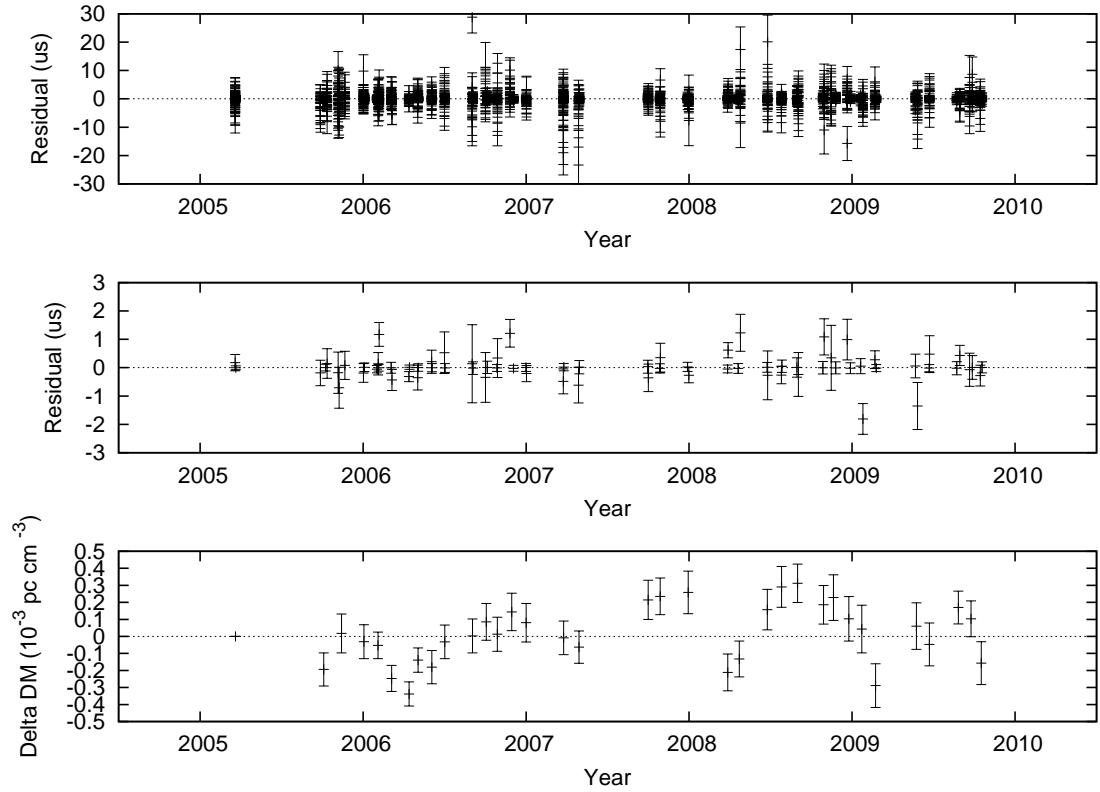}
    \caption{\label{fig:0613:timing}
    Timing summary for PSR~J0613$-$0200, see Figure~\ref{fig:1713:timing} for details.
    }
    \end{center}
    \end{figure*}

    \begin{figure*}[p]
    \begin{center}
    \plotone{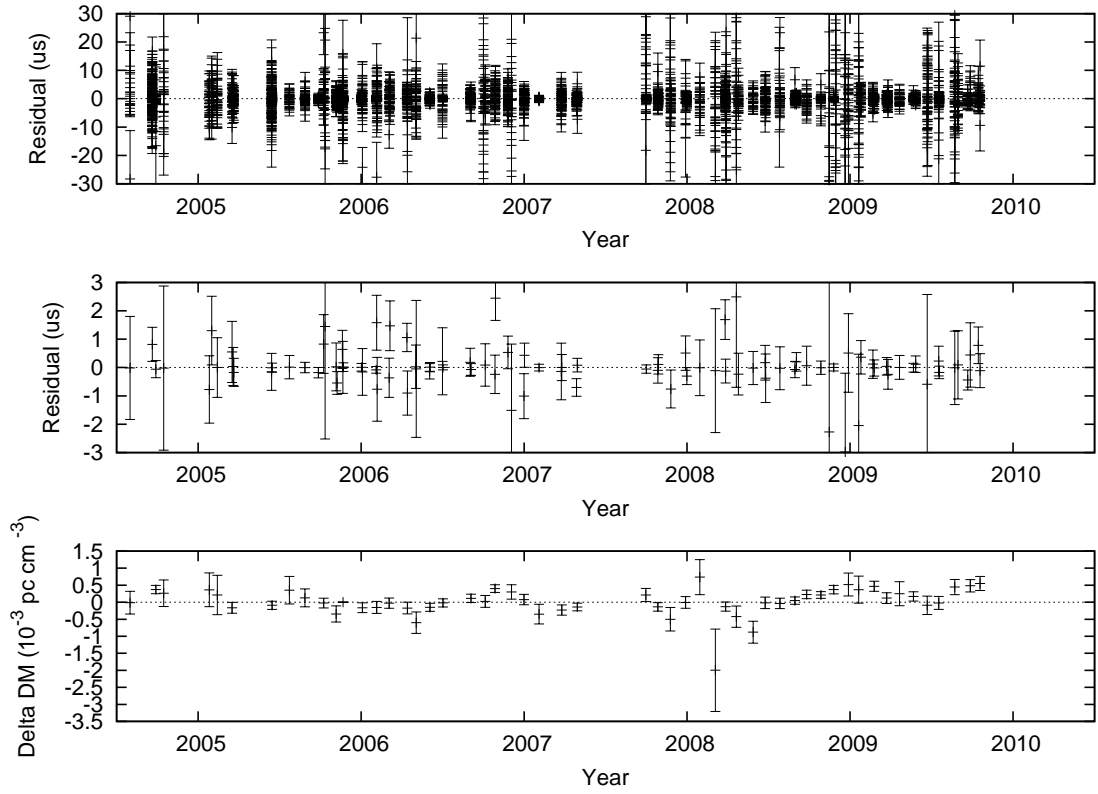}
    \caption{\label{fig:1012:timing}
    Timing summary for PSR~J1012$+$5307, see Figure~\ref{fig:1713:timing} for details.
    }
    \end{center}
    \end{figure*}

    \begin{figure*}[p]
    \begin{center}
    \plotone{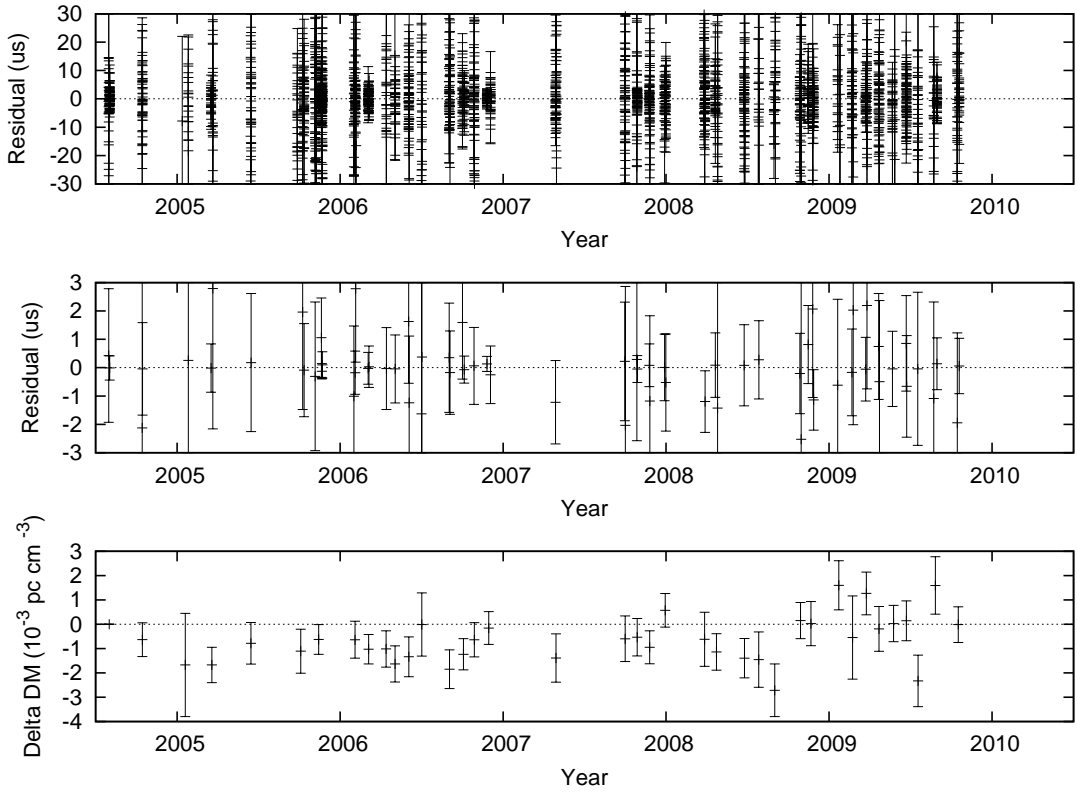}
    \caption{\label{fig:1455:timing}
    Timing summary for PSR~J1455$-$3330, see Figure~\ref{fig:1713:timing} for details.
    }
    \end{center}
    \end{figure*}

    \begin{figure*}[p]
    \begin{center}
    \plotone{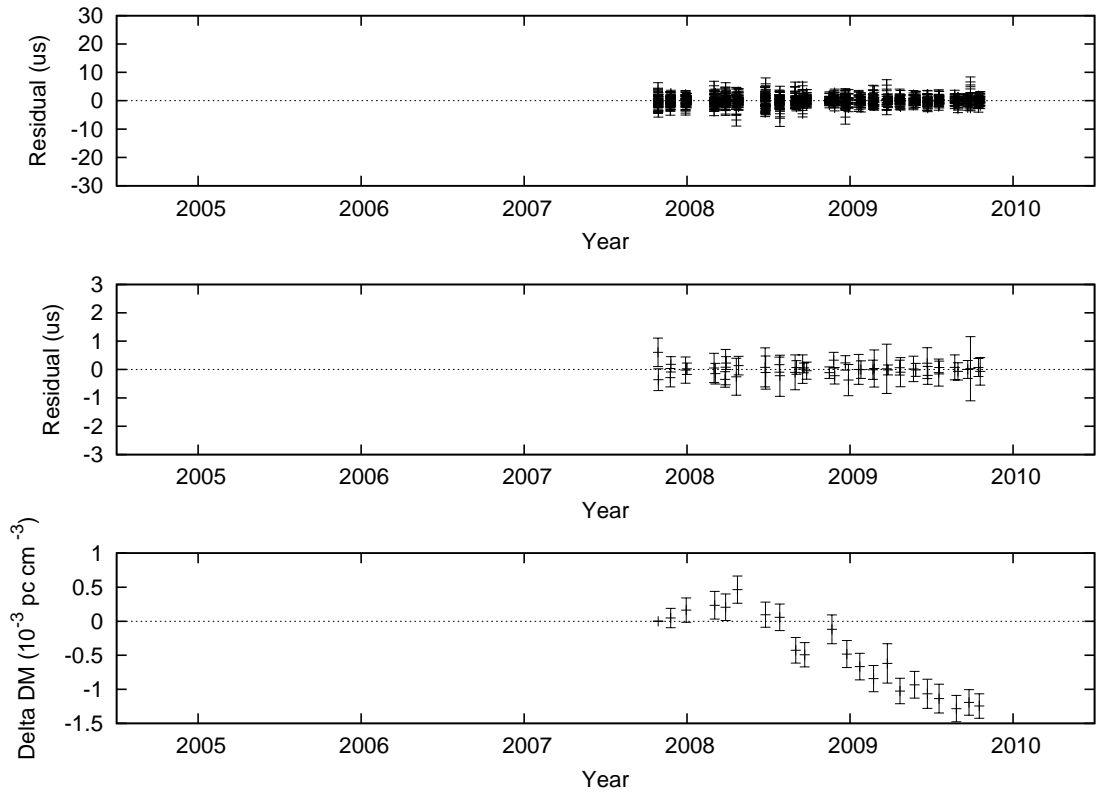}
    \caption{\label{fig:1600:timing}
    Timing summary for PSR~J1600$-$3053, see Figure~\ref{fig:1713:timing} for details.
    }
    \end{center}
    \end{figure*}

    \begin{figure*}[p]
    \begin{center}
    \plotone{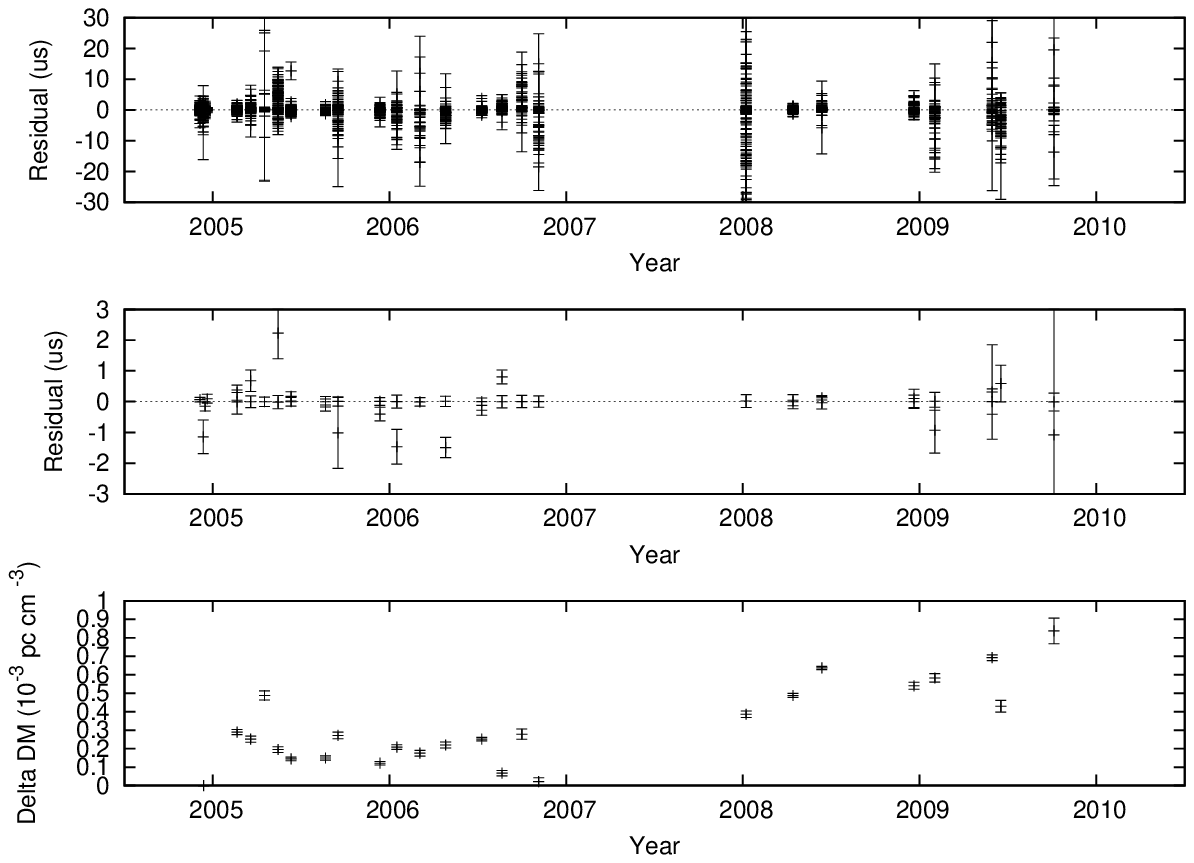}
    \caption{\label{fig:1640:timing}
    Timing summary for PSR~J1640$+$2224, see Figure~\ref{fig:1713:timing} for details.
    }
    \end{center}
    \end{figure*}

    \begin{figure*}[p]
    \begin{center}
    \plotone{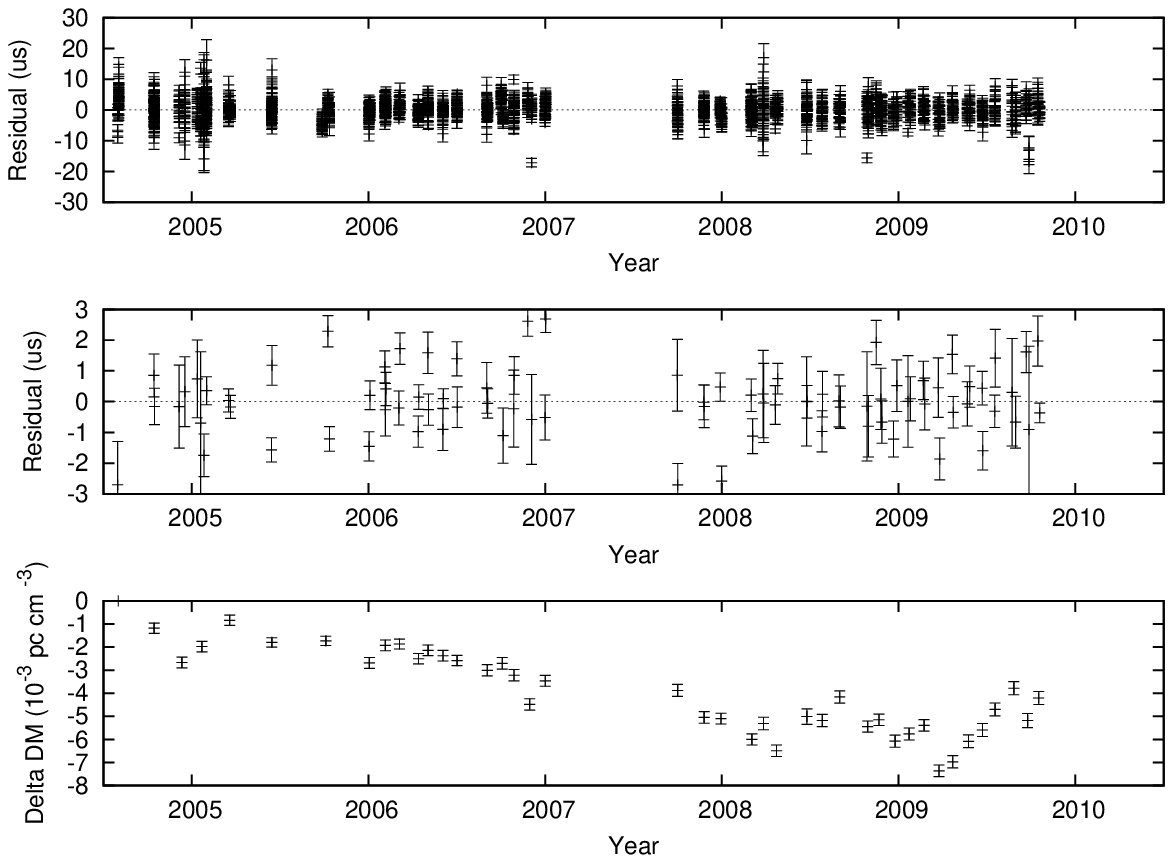}
    \caption{\label{fig:1643:timing}
    Timing summary for PSR~J1643$-$1224, see Figure~\ref{fig:1713:timing} for details.
    }
    \end{center}
    \end{figure*}

    \begin{figure*}[p]
    \begin{center}
    \plotone{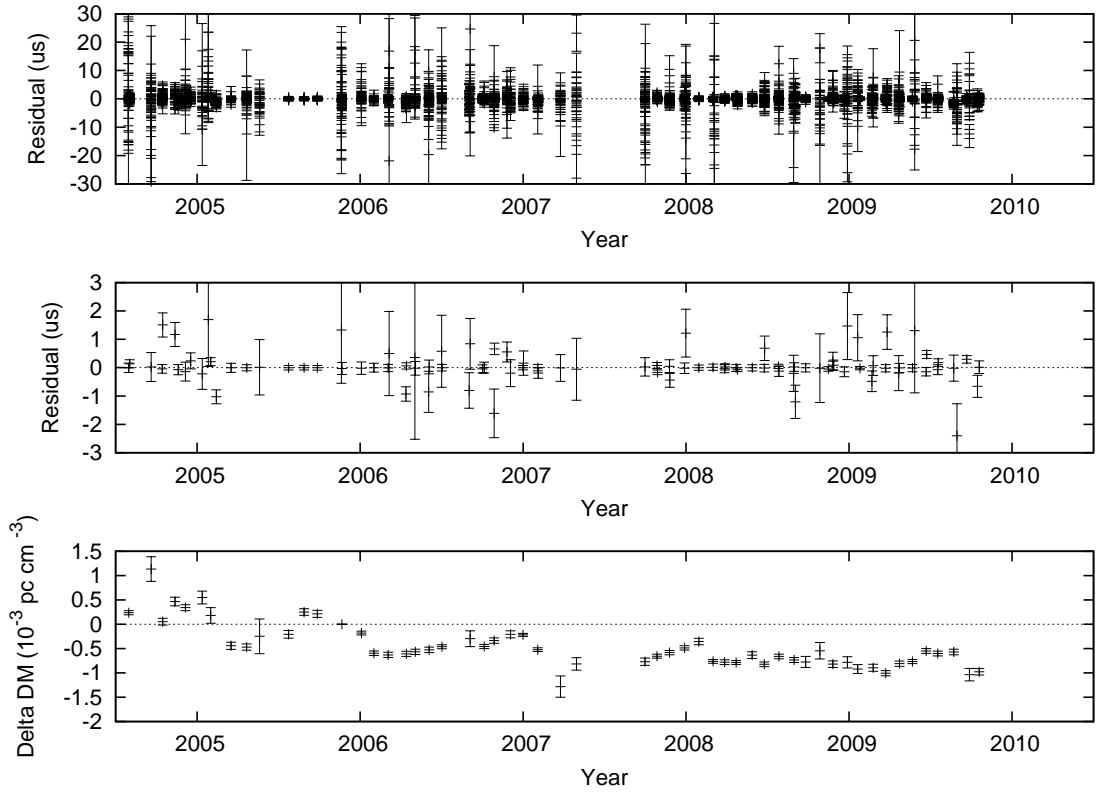}
    \caption{\label{fig:1744:timing}
    Timing summary for PSR~J1744$-$1134, see Figure~\ref{fig:1713:timing} for details.
    }
    \end{center}
    \end{figure*}

    \begin{figure*}[p]
    \begin{center}
    \plotone{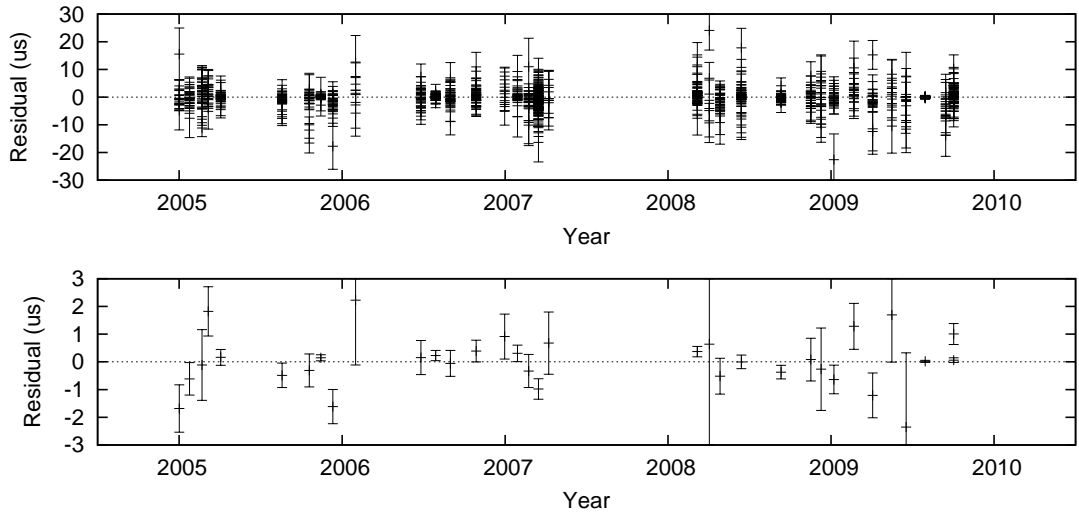}
    \caption{\label{fig:1853:timing}
    Timing summary for PSR~J1853$+$1308, see Figure~\ref{fig:1713:timing} for details.
    }
    \end{center}
    \end{figure*}

    \begin{figure*}[p]
    \begin{center}
    \plotone{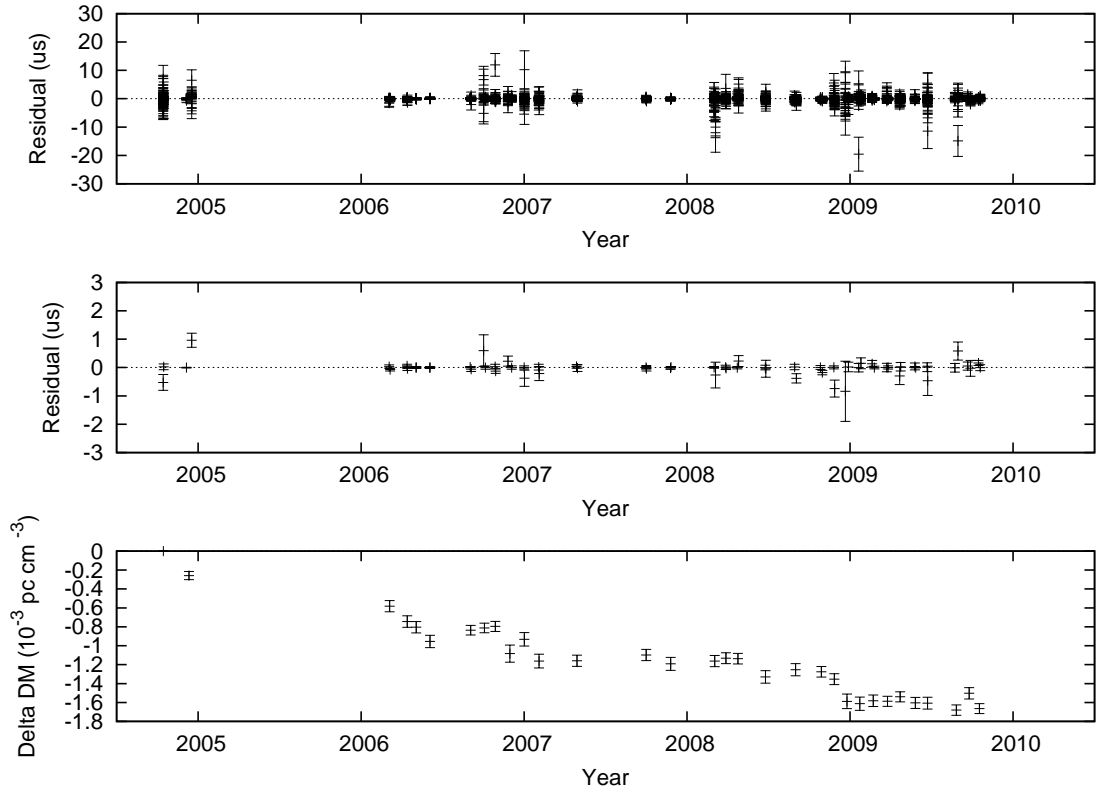}
    \caption{\label{fig:1909:timing}
    Timing summary for PSR~J1909$-$3744, see Figure~\ref{fig:1713:timing} for details.
    }
    \end{center}
    \end{figure*}

    \begin{figure*}[p]
    \begin{center}
    \plotone{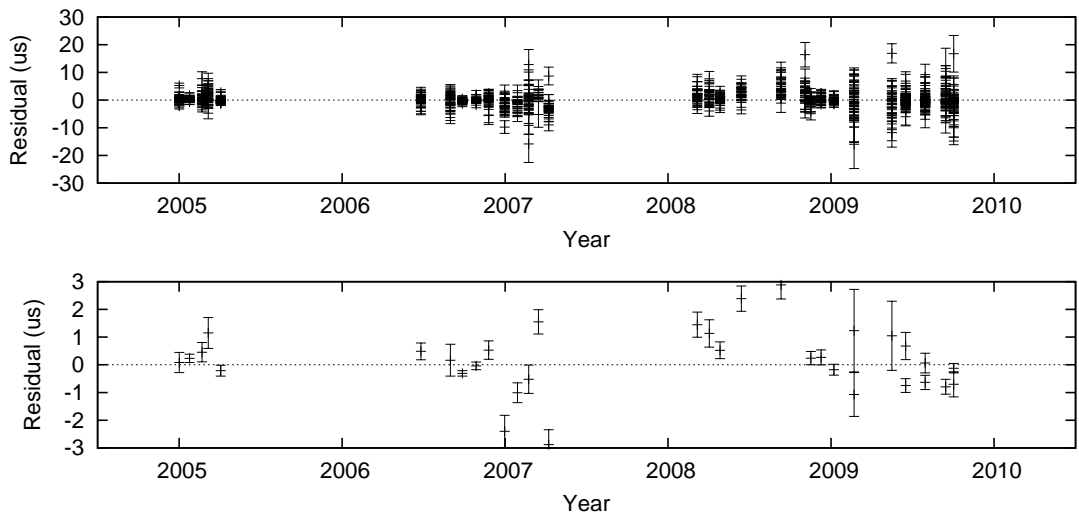}
    \caption{\label{fig:1910:timing}
    Timing summary for PSR~J1910$+$1256, see Figure~\ref{fig:1713:timing} for details.
    }
    \end{center}
    \end{figure*}

    \begin{figure*}[p]
    \begin{center}
    \plotone{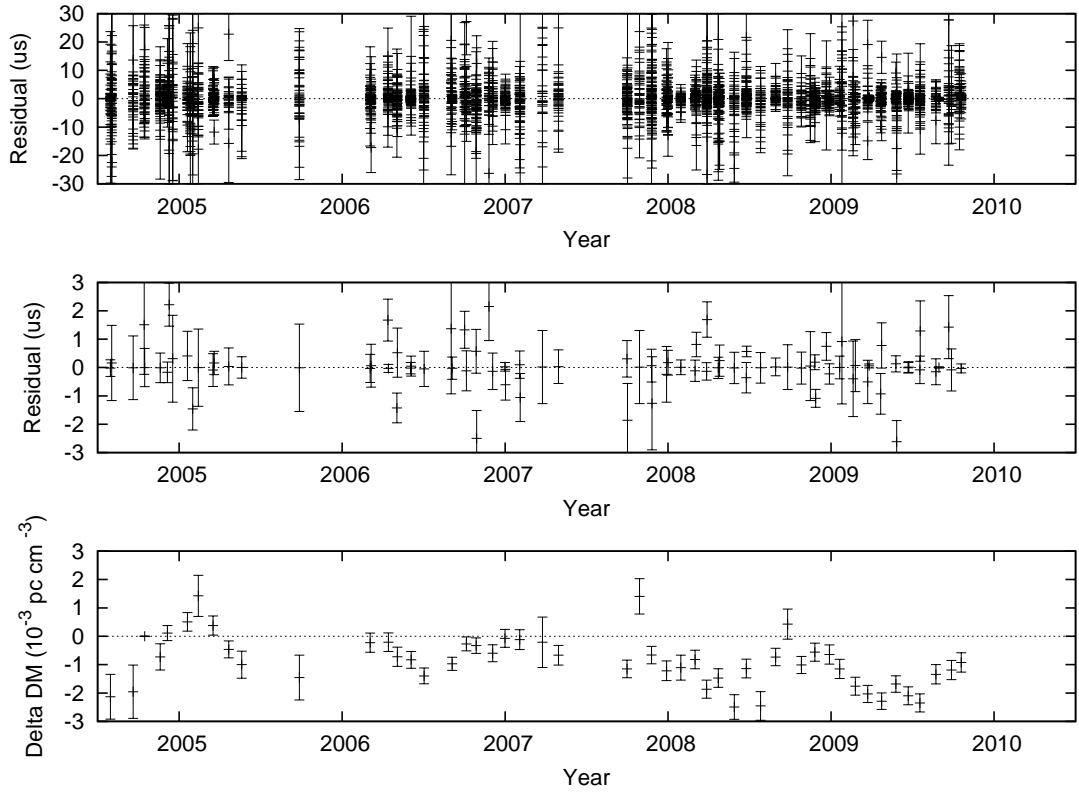}
    \caption{\label{fig:1918:timing}
    Timing summary for PSR~J1918$-$0642, see Figure~\ref{fig:1713:timing} for details.
    }
    \end{center}
    \end{figure*}

    \begin{figure*}[p]
    \begin{center}
    \plotone{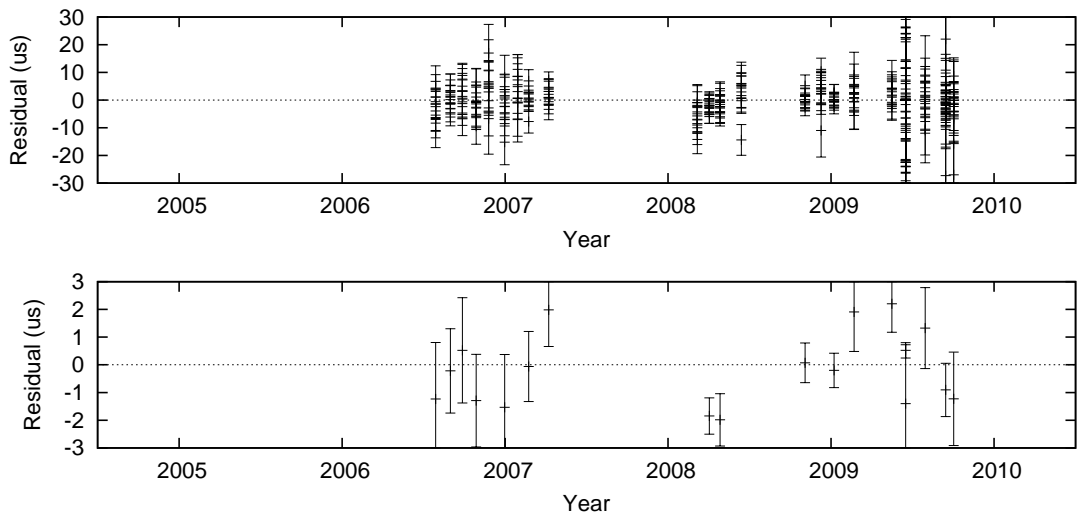}
    \caption{\label{fig:1953:timing}
    Timing summary for PSR~B1953$+$29, see Figure~\ref{fig:1713:timing} for details.
    }
    \end{center}
    \end{figure*}

    \begin{figure*}[p]
    \begin{center}
    \plotone{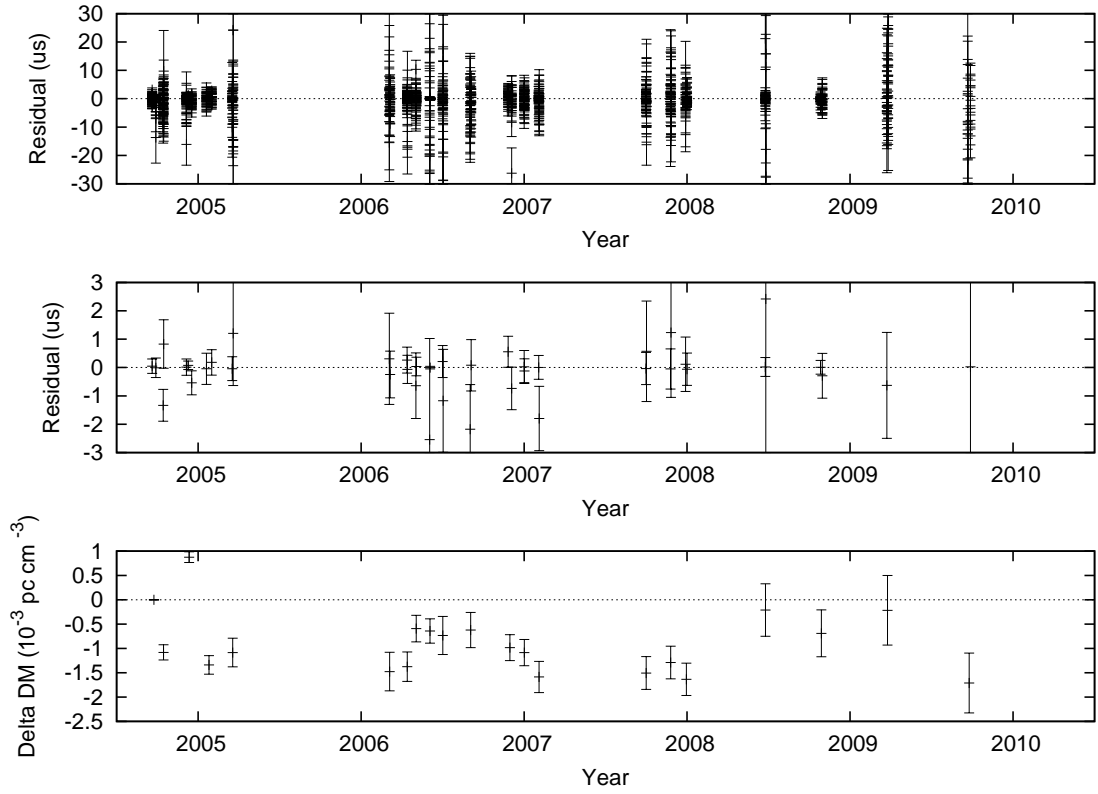}
    \caption{\label{fig:2145:timing}
    Timing summary for PSR~J2145$-$0750, see Figure~\ref{fig:1713:timing} for details.
    }
    \end{center}
    \end{figure*}

    \begin{figure*}[p]
    \begin{center}
    \plotone{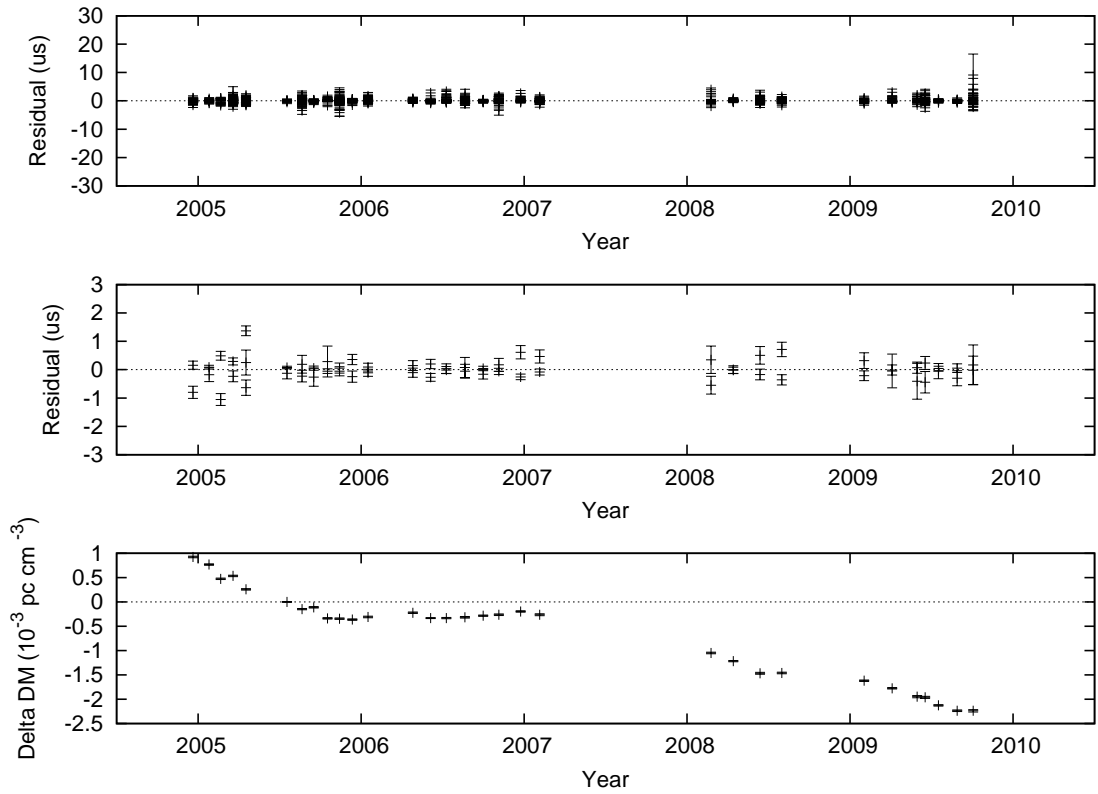}
    \caption{\label{fig:2317:timing}
    Timing summary for PSR~J2317$+$1439, see Figure~\ref{fig:1713:timing} for details.
    }
    \end{center}
    \end{figure*}

\end{document}